\documentclass[twocolumn,showpacs,amsfonts,aps,prc,superscriptaddress,nofootinbib,floatfix]{revtex4-1}
\usepackage{amsfonts}
\usepackage{amsthm}
\usepackage{bm}
\usepackage{graphicx}
\usepackage{amsmath}
\usepackage[usenames,dvipsnames]{color}
\usepackage{soul}
\usepackage{hyperref}

\def\l{\left}
\def\r{\right}

\def\short{\!\!\!}

\def\eq{{\,=\,}}

\newcommand{\ch}{\mbox{ch}}

\newcommand{\vecperp}[1]{{\bm{ #1}}_\perp}
\newcommand{\scaperp}[1]{{ #1_\perp}}

\begin{document}
\setstcolor{red}


\title{Pre-equilibrium evolution effects on heavy-ion collision observables}

\author{Jia Liu}
\email[Correspond to\ ]{liu.2053@osu.edu}
\affiliation{Department of Physics, The Ohio State University,
  Columbus, OH 43210-1117, USA}
\author{Chun Shen}
\affiliation{Department of Physics, McGill University, 3600 University Street, Montreal, Quebec, H3A 2T8, Canada}
\affiliation{Department of Physics, The Ohio State University,
  Columbus, OH 43210-1117, USA}
\author{Ulrich Heinz}
\affiliation{Department of Physics, The Ohio State University,
  Columbus, OH 43210-1117, USA}

\begin{abstract}
In order to investigate the importance of pre-equilibrium dynamics on relativistic heavy-ion collision observables, we match a highly non-equilibrium early evolution stage, modeled by free-streaming partons generated from the Monte Carlo Kharzeev-Levin-Nardi (MC-KLN) and Monte Carlo Glauber (MC-Glb) models, to a locally approximately thermalized later evolution stage described by viscous hydrodynamics, and study the dependence of final hadronic transverse momentum distributions, in particular their underlying radial and anisotropic flows, on the switching time between these stages. Performing a 3-parameter fit of the measured values for the average transverse momenta  $\l< \scaperp{p} \r>$ for pions, kaons and protons as well as the elliptic and triangular flows of charged hadrons $v_{2,3}^\mathrm{ch}$, with the switching time $\tau_s$, the specific shear viscosity $\eta/s$ during the hydrodynamic stage, and the kinetic decoupling temperature $T_\mathrm{dec}$ as free parameters, we find that the preferred ``thermalization'' times $\tau_s$ depend strongly on the model of the initial conditions. MC-KLN initial conditions require an earlier transition to hydrodynamic behavior (at $\tau_s \approx$ 0.13 fm/$c$)  , followed by hydrodynamic evolution with a larger specific shear viscosity $\eta/s\approx$ 0.2, than MC-Glb initial conditions which prefer switching at a later time ($\tau_s\approx$ 0.6 fm/$c$) followed by a less viscous hydrodynamic evolution with $\eta/s\approx$ 0.16. These new results including pre-equilibrium evolution are compared to fits without a pre-equilbrium stage where all dynamic evolution before the onset of hydrodynamic behavior is ignored. In each case, the quality of the dynamical descriptions for the optimized parameter sets, as well as the observables which show the strongest constraining power for the thermalization time, are discussed.
 \end{abstract}


\date{\today}

\maketitle

%

\section{Introduction}
\label{sec1}

The collision of two fast moving heavy nuclei produces a new form of matter, quark-gluon plasma (QGP). It consists of deconfined quarks and gluons, and in the early years QGP was thought to behave like a weakly-interacting gas. However, experiments conducted at the Relativistic Heavy Ion Collider (RHIC) and Large Hadron Collider (LHC) show strong collective flow, revealing that the QGP is in fact a strongly coupled liquid, with very low viscosity. The evolution of this strongly coupled QGP is well described by hydrodynamics, which successfully reproduced and even predicted the experimentally observed transverse momentum spectra and flow anisotropies of emitted hadrons. 

The validity of hydrodynamics relies on the matter being close to local thermal equilibrium. Within a purely hydrodynamic approach that ignores details about how the system approaches equilibrium, the observed large hadron momentum anisotropy can only be explained if thermalization happens fast and the hydrodynamic expansion does not begin later than about 1fm/$c$ after the two nuclei impact each other \cite{Kolb:2000sd}. This finding calls for a mechanism to explain how the matter produced in the collision can thermalize so fast. Recently, significant work has been done on modeling the thermalization process during the pre-equilibrium stage \cite{Broniowski:2008qk, Gelis:2013rba,  Wu:2011yd, Romatschke:2013re,Ryblewski:2013eja, Heller:2012km}. While these studies have not yet fully explained the rapid thermalization, some of them have indicated that pre-equilibrium evolution can have a non-negligible influence on the final observables. 

In this paper, we return to the question of how quickly thermalization must happen for a hydrodynamic approach to provide a successful description of experimental data, by studying the weakly interacting limit of pre-equilibrium dynamics. Freely streaming partonic quanta can be considered as the extreme limit of a weakly interacting system and the diametrically opposite to a fluid dynamical description which requires strong coupling. By coupling a free-streaming pre-equilibrium stage to hydrodynamics and varying the switching time $\tau_s$ one can smoothly interpolate between a very strongly coupled (small $\tau_s$, early transition to hydrodynamics) and weakly coupled (large $\tau_s$, late transition to hydrodynamic behavior) early evolution stage. We are interested in finding the largest $\tau_s$ value that is compatible with phenomenology. Replacing the non-interacting free-streaming parton stage in our model by a model in which the matter constituents interact more realistically with modest interaction strength should allow for an earlier transition to hydrodynamic behavior. For this reason we expect our approach to yield a robust upper limit, for a fixed viscosity during the subsequent hydrodynamic stage, for the time at which the matter created in the collision must have reached a sufficient degree of local equilibrium to be described hydrodynamically. The dependence of this upper limit on the viscosity in the hydrodynamic stage will also be explored.

It was initially thought that substituting the earliest stage in a hydrodynamic evolution model by a free-streaming gas would delay the buildup of thermodynamic pressure and thus reduce the finally observed collective flow \cite{Kolb:2000sd}. This ignored, however, the fact that in a spatially inhomogeneous medium free-streaming (or, for that matter, any kind of pre-equilibrium evolution) generates strong position-momentum correlations which, upon thermalization, lead to strong initial flow in the hydrodynamic stage. We show here that this ``pre-flow'' actually increases the finally observed radial flow, with consequences that are opposite to the expectations reported in Refs.~\cite{Kolb:2000sd}. Here we explore the evolution of the widely implemented MC-KLN \cite{Kharzeev:2001yq, Kharzeev:2004if} and MC-Glauber \cite{Alver:2008aq} initial conditions at LHC energies, allowing them to be evolved by free-streaming for a time $\tau_s$ before switching to a viscous hydrodynamic description for the rest of the evolution.

Our work focuses on the effects brought by free-streaming on both the hydrodynamic initial conditions and the final observables. We mainly focus on the MC-KLN model which provides a complete prediction for the initial gluon distribution, not only in space but also in momentum. However, we show that for massless partons moving with the speed of light the shape of the initial momentum distribution is irrelevant as long as it is locally isotropic. This allows to apply our description also to MC-Glb initial conditions although that model makes no prediction per se about the initial parton momentum distribution. Free-streaming evolves the initial conditions from an initial parton formation time $\tau_0$ (which is taken to be very close to zero) to the switching time $\tau_s$ when we switch  to a near-equilibrium hydrodynamic description. The sudden transition to approximate local equilibrium is implemented by applying the Landau matching procedure. By tuning the switching time, we can enforce fast thermalization by setting $\tau_s{\,\simeq\,}\tau_0$, or slow thermalization by setting $\tau_s{\,\gg\,}\tau_0$. The hydrodynamic initial conditions obtained from the Landau matching procedure vary with the switching time, enabling an investigation of the influence of $\tau_s$ on the final observables. The hydrodynamic evolution is performed with the code VISH2+1 \cite{Song:2007ux, Shen:2014vra}, without hadron cascade afterburner. For the hydrodynamic evolution, we use a constant specific shear viscosity $\eta/s$. 
Freeze-out is implemented at  a fixed kinetic freeze-out temperature $T_\mathrm{dec}$, followed by a Cooper-Frye procedure with full resonance decay cascade to convert the hydrodynamic output into final stable particle spectra. 

In Section \ref{fsFormalism}, we outline the free-streaming evolution in the pre-equilibrium stage and describe the Landau matching procedure. Its consequences on the hydrodynamic initial conditions are discussed in Section \ref{fsEffectsonICs}. Section \ref{fsEffects} shows how the hydrodynamical evolution responds to initial conditions generated at different switching times. A difficulty related to the conversion of partons to hadrons that arises from a late switching time $\tau_s$ is discussed and resolved in Section \ref{fsIssues}. In Sections \ref{energyFlow} and \ref{upperLimitOfFS}, the energy flow anisotropy and hadron mean transverse momenta are constructed to illustrate how final observables change with switching time. Finally in Section \ref{parameterSearch} we introduce a multidimensional parameter search procedure to systematically study the preferred ranges of $\tau_s$, $\eta/s$ and $T_\mathrm{dec}$. For both MC-KLN and MC-Glb initial conditions, both with and without a free-streaming pre-equilibrium stage before the onset of hydrodynamic behavior, we determine the best-fit parameters and their uncertainty ranges, and discuss their relative quality of describing the data. Conclusions are presented in Section \ref{paperConclusion}.

\section{Formulation of free-streaming and Landau matching}
\label{fsFormalism}

The evolution of partons in the free-streaming model is described by the collisionless Boltzmann equation
\begin{eqnarray}
p^\mu\partial_\mu f(x,p)=0.
\label{collisionlessBoltzmann}
\end{eqnarray}
We work in Milne coordinates and write $f(x, p) = f(\bm{x}_\perp, \eta_s, \tau;\, \bm{p}_\perp, y)$,
with longitudinal proper time $\tau\eq\sqrt{t^2{-}z^2}$, space-time rapidity $\eta_s\eq\frac{1}{2}\ln[(t{+}z)/(t{-}z)]$, and rapidity $y\eq\frac{1}{2}\ln[(E{+}p_z)/(E{-}p_z)]$. We assume massless partons for which $E\eq|\bm{p}|\eq\sqrt{\vecperp{p}^2{+}p_z^2}$. The collision\-less Boltzmann equation is easily solved analytically, relating the final parton distribution $f(\bm{x}_\perp, \eta_s, \tau_s; \bm{p}_\perp, y)$ to the $f(\bm{x}_\perp, \eta_s, \tau_0; \bm{p}_\perp, y)$ by a spatial coordinate shift, keeping the $\vecperp{p}$ distribution unchanged. For massless partons one finds:
\begin{eqnarray}
f(\bm{x}_\perp, \eta_s, \tau_s; \bm{p}_\perp, y)%
{\eq}f(\bm{x}_\perp{-}(\tau_s{-}\tau_0)\hat{\bm{p}}_\perp, \eta_s, \tau_0; \bm{p}_\perp, y),\quad\
\label{fsSolution}
\end{eqnarray}
where $\hat{\bm{p}}_\perp\eq\vecperp{p}/\scaperp{p}\eq(\cos\phi_p, \sin\phi_p)$, with $\phi_p$ being the azimuthal angle of $\vecperp{p}$ in the plane transverse to the beam. The initial distribution $f(\bm{x}_\perp, \eta_s, \tau_0; \bm{p}_\perp, y)$ is assumed to be locally isotropic in transverse momentum, i.e. independent of $\phi_p$: $f_0=f(\bm{x}_\perp, \eta_s, \tau_0; p_\perp, y)$.

To initialize the hydrodynamic we must decompose the energy momentum tensor $T^{\mu\nu}$ in hydrodynamic form. In the free-streaming stage $T^{\mu\nu}(x)$ can be obtained from the solution of Eq.~(\ref{fsSolution}) for the parton distribution as  
\begin{eqnarray}
T^{\mu\nu}({\bm x}_\perp, \eta_s, \tau_s)\eq\frac{g}{(2\pi)^3}\! \int\!
\frac{d^3p}{E} p^\mu p^\nu f(\bm{x}_\perp, \eta_s, \tau_s; \bm{p}_\perp, y),\quad\
\label{TmunuDef}
\end{eqnarray}
where for massless particles $p^0\eq{E}\eq|{\bm p}|$ and $g$ is a degeneracy factor. We assume longitudinal boost-invariance and restrict the dependence of $f(x, p)$ on $\eta_s$ and $y$ as follows \cite{Bjorken:1982qr, Baym:1984np}:
\begin{eqnarray}
f({\bm x_\perp}, \eta_s, \tau;\, \vecperp{p}, y)
= \frac{\delta(y{-}\eta_s)}{\tau m_\perp \ch(y{-}\eta_s)}
\tilde{f}(\vecperp{x}, \tau;\, \vecperp{p}, y).\quad
\label{boostInvariantF}
\end{eqnarray}
Again, for massless partons $m_\perp=\sqrt{m^2+\vecperp{p}^2}=\scaperp{p}$. For a boost-invariant system it is sufficient to know $T^{\mu\nu}$ in the transverse plane at $z=0$ (i.e. $\eta_s$=0) where 
\begin{eqnarray}
T^{\mu\nu}({\bm x}_\perp, \eta_s{=}0, \tau)
&=& \frac{g}{(2\pi)^3} \frac{1}{\tau}\int_{0}^{\infty} d\scaperp{p} \int_{-\pi}^{\pi} d\phi_p 
\nonumber\\
&&\times\l.\l[p^\mu p^\nu \tilde{f}(\vecperp{x}, \tau;\, \vecperp{p}, y) \r]\r|_{y=0}.
\label{TmunuboostInvariant}
\end{eqnarray}
The factor $1/\tau$ is characteristic for systems with boost-invariant longitudinal expansion.

Inspection of this formula shows that for massless partons the evolution of $T^{\mu\nu}(\bm{x}_\perp, \eta_s{=}0, \tau)$ does not depend on the underlying parton momentum distribution. For massless partons that are initially distributed locally isotropically in $\vecperp{p}$ the spatial distribution of the energy momentum tensor $T^{\mu\nu}(\vecperp{x}, \tau)$ at time $\tau$ depends only on its initial spatial distribution at time $\tau_0$, but not on the $\scaperp{p}$-distribution (which may or may not depend on $\vecperp{x}$). One sees this by observing that the integral on the right hand side of Eq.~(\ref{TmunuboostInvariant}) can be rewritten as
\begin{eqnarray}
&\;&\int_{0}^{\infty} d\scaperp{p} \int_{-\pi}^{\pi} d\phi_p 
\l.\l[p^\mu p^\nu \tilde{f}(\vecperp{x}, \tau;\, \vecperp{p}, y) \r]\r|_{y=0} \nonumber \\ 
&\short = & \int_0^\infty \!\! p_\perp^2 dp_\perp  
                   \int_{-\pi}^{\pi} \!\! d\phi_p\; \hat{p}^\mu \hat{p}^\nu 
                   \tilde{f}(\vecperp{x}{-}(\tau{-}\tau_0)\vecperp{\hat{p}}, \tau_0; p_\perp, 0)\quad
\label{TmunuIntegration}
\end{eqnarray}
where $\hat{p}^\mu \equiv \l.\frac{p^\mu}{\scaperp{p}}\r|_{y=0}$ depends only on $\phi_p$. Hence 
\begin{eqnarray}
T^{\mu\nu}(\bm{x}_\perp, \eta_s{=}0, \tau) = \frac{1}{\tau}\int_{-\pi}^{\pi}d\phi_p\; \hat{p}^\mu \hat{p}^\nu F(\bm{x}_\perp, \tau; \phi_p),
\label{TmunuSimplified_outer}
\end{eqnarray}
where 
\begin{eqnarray}
&& F({\bm x_\perp}, \tau; \phi_p) = F_0(\vecperp{x}{-}(\tau{-}\tau_0)\vecperp{\hat{p}})
\nonumber\\
&&\equiv \frac{g}{(2\pi)^3} \int_{0}^{\infty} p_\perp^2 dp_\perp 
        \tilde{f}(\vecperp{x}{-}(\tau{-}\tau_0)\vecperp{\hat{p}}, \tau_0; \scaperp{p}, 0)
 \label{TmunuSimplified_innner}
\end{eqnarray}
is independent of how $f_0$ depends on the magnitude of $\scaperp{p}$, and $F_0(\vecperp{x})$ denotes the spatial distribution function (integrated over momenta) at $\tau\eq\tau_0$.

At the switching time $\tau_s$, the solution (\ref{TmunuSimplified_outer}) for $T^{\mu\nu}$ is decomposed in viscous hydrodynamic form:
\begin{eqnarray}
T^{\mu\nu} = e u^\mu u^\nu -(\mathcal{P}+\Pi) \Delta^{\mu\nu} + \pi^{\mu\nu}.
\label{TmunuHydroForm}
\end{eqnarray}
Here $e$ and $\mathcal{P}$ are the energy density and pressure in the local fluid rest frame (LRF), $\Pi$ is the local bulk viscous pressure, $\pi^{\mu\nu}$ is the shear pressure tensor, and $u^\mu$ is the local fluid velocity. The projection operator $\Delta^{\mu\nu}\equiv g^{\mu\nu}{-}u^\mu u^\nu$ projects on the spatial coordinates in the LRF, and the spacetime metric $g^{\mu\nu}$ in Milne coordinates is given by $g^{\mu\nu}=\mathrm{diag}(1,-1,-1,-1/\tau^2)$.

The Landau matching condition defines the fluid rest frame velocity as the time-like eigenvector of $T^{\mu\nu}$, and the energy density $e$ as its eigenvalue:
\begin{eqnarray}
T^{\mu\nu} u_{\nu} = e u^\mu,
\label{LandauMatchingConditions}
\end{eqnarray}
with $u^\mu u_{\mu} = 1$. If such a solution exists \cite{Arnold:2014jva} it is unique since $T^{\mu\nu}$ has at most one time-like eigenvector. 

Landau matching conserves the system's total energy, but not its entropy. Due to the absence of collisions, Eq.~(\ref{collisionlessBoltzmann}) conserves entropy. The solution (\ref{fsSolution}) therefore yields the same entropy at $\tau_s$ and $\tau_0$. After matching, however, the entropy density $s=\partial\mathcal{P}/\partial T$ is related to the energy density $e$ and pressure $\mathcal{P}$ by the thermalized fluid's equation of state (EOS) $\mathcal{P}=\mathcal{P}(e)$. This implies that in general the total entropy $\mathcal{S}$ of the system increases discontinuously at the switching time $\tau_s$. This sudden increase is the consequence of the assumed sudden thermalization of the system that is implicit in the Landau matching procedure. For successful phenomenology, we have to normalize the entropy density profile after Landau matching such that, upon completion of the dynamical evolution, it correctly reproduces the observed final multiplicity $dN_{ch}/dy$. This is done for central collisions, and the predicted impact parameter dependence of the final $dN_{ch}/dy$ is then taken as an argument for or against the validity of the initial model used to generate the initial conditions. In spite of the entropy jump at $\tau_s$, the normalization of the entropy density profile after Landau matching has a one-to-one relation with the normalization of the initial distribution function. Since the entropy jump depends on the chosen value of the switching time, this initial normalization also depends on $\tau_s$: to preserve the same $dN_{ch}/dy$ at the end of the hydrodynamic evolution thus requires a renormalization of the initial distribution function when $\tau_s$ is varied. 

After finding the energy density, the thermodynamic pressure is given by the EOS $\mathcal{P} = \mathcal{P}(e)$ of the thermalized liquid. The dynamically induced bulk viscous pressure is then reconstructed from $T^{\mu\nu}$ by using the identity
\begin{eqnarray}
\Pi = -\frac{1}{3}  {\textrm{Tr}} (\Delta_{\mu\nu}T^{\mu\nu}) - \mathcal{P}.
\label{bulkPi}
\end{eqnarray}
Finally, the shear pressure tensor is obtained by contracting $T^{\mu\nu}$ with the double projection operator $\Delta^{\mu\nu}_{\alpha\beta}\equiv\frac{1}{2}\l( \Delta^\mu_\alpha \Delta^\nu_\beta{+}\Delta^\mu_\beta \Delta^\nu_\alpha\r) - \frac{1}{3}\Delta^{\mu\nu}\Delta_{\alpha\beta}$,
\begin{eqnarray}
\pi^{\mu\nu} = \Delta^{\mu\nu}_{\alpha\beta} T^{\alpha\beta}.
\label{pimnExtraction}
\end{eqnarray}
Alternatively, 
\begin{eqnarray}
\pi^{\mu\nu} = T^{\mu\nu}-e u^\mu u^\nu +(\mathcal{P}+\Pi)\Delta^{\mu\nu},
\label{pimnExtraction_bySubtraction}
\end{eqnarray}
where $T^{\mu\nu}$ is from Eq.~(\ref{TmunuSimplified_outer}), $u^\mu$ and $e$ from Eq.~(\ref{LandauMatchingConditions}), $\mathcal{P}$ from EOS and $\Pi$ from Eq.~(\ref{bulkPi}). 

%
\begin{figure*}[!htb]
    \includegraphics[width=0.90\linewidth]{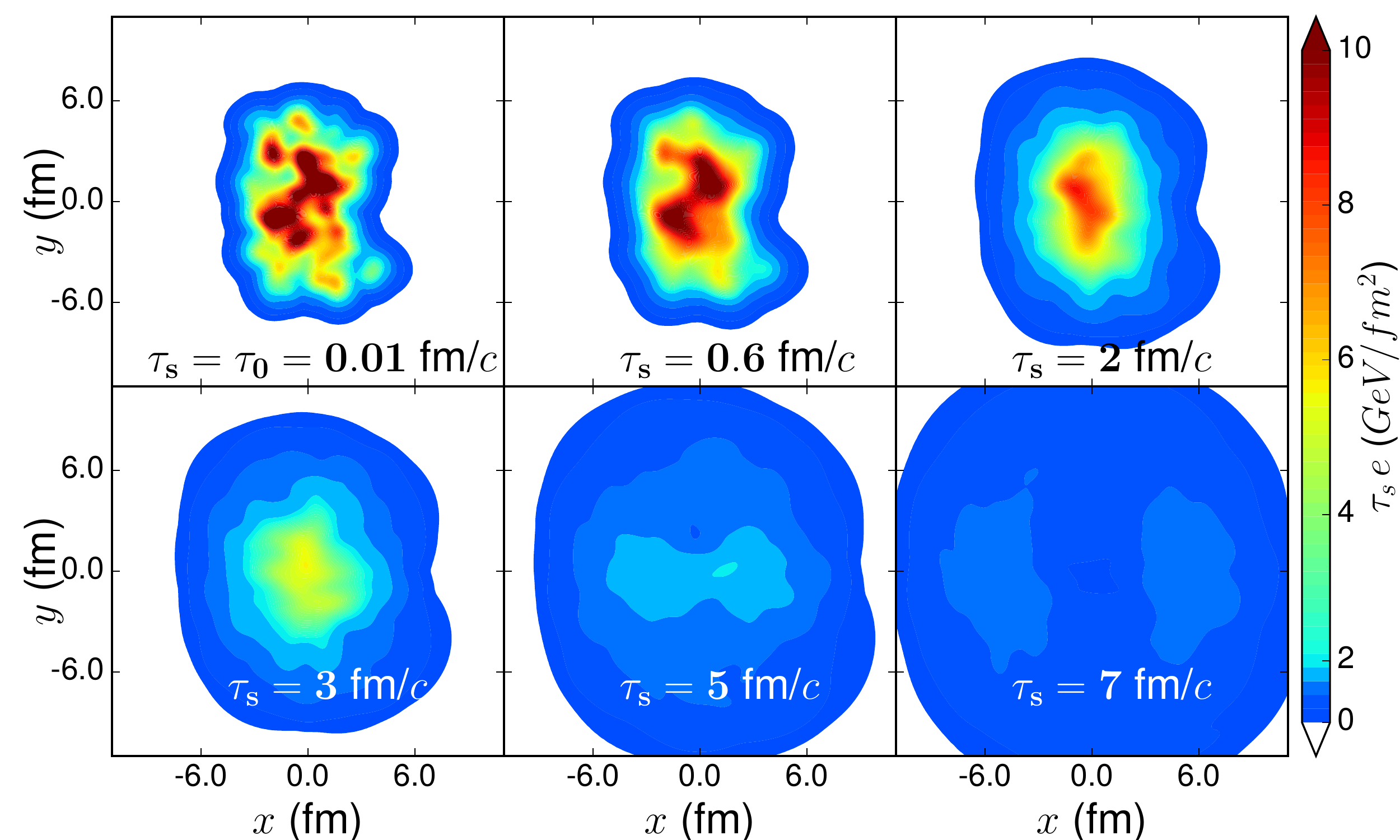}
    \caption{(Color online)
    Contour plots of the local energy density multiplied by $\tau_s$ in the transverse 
    plane, for a single fluctuating Pb+Pb collision event  of 10\% -- 20\% centrality at 
    $\sqrt{s}=2.76$ $A$TeV. To show the edge of the fireball more clearly, white is 
    used for $\tau_s e$ below $10^{-4}\,\text{GeV/fm}^2$, and dark red is used for 
    $\tau_s e >10\,\text{GeV/fm}^2$.
    \label{F1}
}
\end{figure*}
%

The initial conditions for $u^\mu$, $e$, $\mathcal{P}$, $\Pi$ and $\pi^{\mu\nu}$ obtained from Eqs.~(\ref{LandauMatchingConditions})-(\ref{pimnExtraction}) are more realistic than those used in many recent implementations of hydrodynamics where any evolution of the system of the system between $\tau_0$ and the hydrodynamic starting time $\tau_s$ is ignored and the hydrodynamic stage is initialized with zero transverse flow and an ad hoc guess for the viscous pressure components. We will see that the non-zero flow velocities and well-defined non-zero initial conditions for $\Pi$ and $\pi^{\mu\nu}$ resulting from the Landau matching procedure (\ref{LandauMatchingConditions}) have important consequences for the subsequent hydrodynamic evolution and final observables. In this work, we focus on the effects of shear viscosity on the evolution of the collision systems. We implement our assumption of vanishing bulk viscosity $\zeta$ by setting $\zeta/s =10^{-6}$ and then using the second order evolution equation \cite{Muronga:2001zk} ($D{\,\equiv\,}u^\mu \partial_\mu$)
\begin{equation}
D \Pi = -\frac{1}{\tau_\Pi} (\Pi + \zeta \theta) - \frac{1}{2} \Pi \frac{\zeta T}{\tau_\Pi} \partial_\mu \left(\frac{\tau_\Pi}{\zeta T} u^\mu \right)
\end{equation}
to evolve the bulk viscous pressure $\Pi$ dynamically to zero, on a microscopic time scale given by the bulk relaxation time $\tau_\Pi = \frac{3}{4 \pi T}$ \cite{Song:2009rh}. 

\section{Hydrodynamic initial conditions after a free-streaming pre-equilibrium stage}
\label{fsEffectsonICs}

In this and the following sections where we investigate the qualitative effects of a free-streaming pre-equilibrium stage on the hydrodynamical evolution and final observables,  we focus on MC-KLN initial conditions \cite{Kharzeev:2001yq, Kharzeev:2004if}. We will return to the MC-Glauber model in Sec.~\ref{upperLimitOfFS}. 

Compared to hydrodynamics, free-streaming dilutes the local energy density much faster. Because there are no collisions, signals carried by the massless partons move with the speed of light instead of the smaller drift velocity that would characterize an interacting medium (for a thermalized medium this would be the speed of sound). In Fig.~\ref{F1}, the local energy density just after switching to hydrodynamics is shown at different switching time $\tau_s$. In the first panel, Landau matching is implemented at the matter formation time $\tau_0$ (taken as $\tau_0$=0.01 fm/$c$). At this time, the matter distribution features many hot spots in the transverse plane, reflecting the fluctuating nucleon positions that, through the nuclear thickness function, control the saturation momentum and thus the density of the produced gluons in the MC-KLN model. As the switching time increases, the bumps in the energy density spread and gradually dissolve as a result of the free-streaming of the partons. The initially bumpy energy density profile becomes much smoother and less deformed, resulting in decreasing spatial eccentricity coefficients $\mathcal E_n$.

The spatial eccentricity coefficient of harmonic order $n$ at the beginning of hydrodynamical stage is usually defined as \cite{Alver:2010gr, Alver:2010dn}
\begin{eqnarray}
\mathcal E_n = \epsilon_n e^{in\Phi_n} 
= -\frac{\int d^2r_\perp\, r^n_\perp\, e^{in\phi}\, e(\vecperp{r})}
{\int d^2r_\perp\, r^n_\perp\, e(\vecperp{r})}, \quad n>1,\quad
\label{eccentricityDefBefore}
\end{eqnarray}
where $ e(\vecperp{r})$ is the LRF energy density obtained from Eq.~\ref{LandauMatchingConditions}, and the minus sign ensures that the angle $\Phi_n$ points to the direction where energy density falls fastest. (For fluctuating events, the energy density profile must be re-centered to the origin in the transverse plane before calculating the eccentricities.) However, if the initial conditions feature a non-zero initial hydrodynamic flow profile, as is the case in Eq.~(\ref{LandauMatchingConditions}), flow anisotropies cause the Lorentz contraction factor $\gamma$ between the local and global rest frame to depend on the azimuthal angle $\phi$. In the laboratory frame, the initial energy density is thus better characterized by eccentricity coefficients calculated with a modified prescription using the energy density in the laboratory frame:
\begin{eqnarray}
\label{eccentricityDef}
\mathcal E_n(\tau_s) &=& \epsilon_n(\tau_s) e^{in\Phi_n(\tau_s)}\\
&=&-\frac{\int_{\tau_s} d^3\sigma_\mu (x)\, T^{\mu\nu}(x)\,u_\nu(x)\, r_\perp^n\, e^{in\phi}}
{\int_{\tau_s} d^3\sigma_\mu (x)\, T^{\mu\nu}(x)\,u_\nu(x)\, r_\perp^n} 
\nonumber\\
&=& -\frac{\int d^2r_\perp \gamma(\vecperp{r})\, e(\vecperp{r})\, r_\perp^n\, e^{in\phi} }
{\int d^2r_\perp \gamma(\vecperp{r})\, e(\vecperp{r})\, r^n_\perp}, \quad\quad (n>1) 
\nonumber
\end{eqnarray}
where $d^3\sigma_\mu$ is the normal vector on the switching hypersurface of constant $\tau_s$. Now $\Phi_n$ points in the direction of the steepest descent in the lab frame. If need be, the modified eccentricity definition (\ref{eccentricityDef}) can also be used for different switching surfaces.

\begin{figure}[!hbt]
    \includegraphics[width=\linewidth]{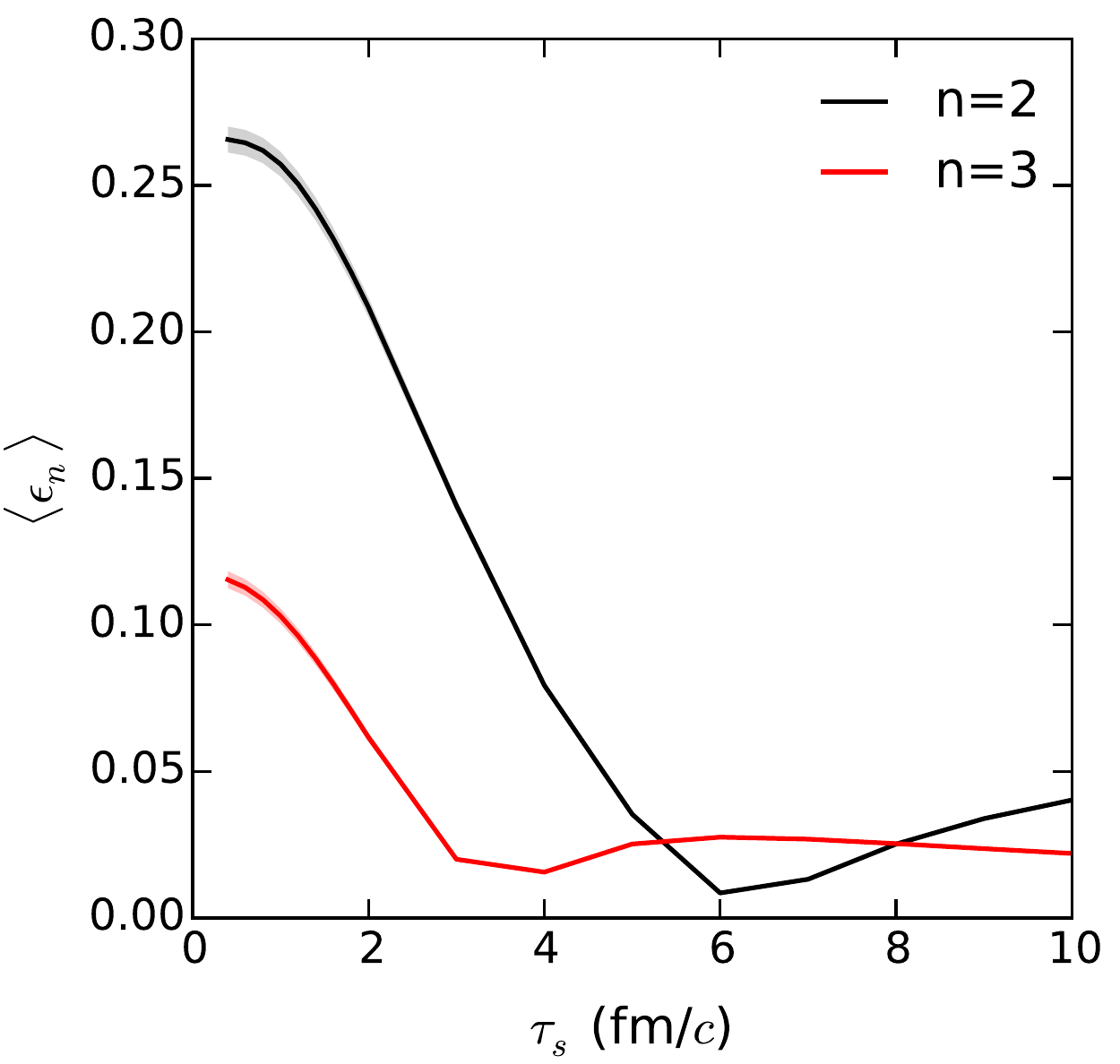}
    \caption{(Color online)
    Event-averaged spatial eccentricity as a function of switching time for an 
    ensemble of 400 fluctuating Pb+Pb collision events of 10\% -- 20\% centrality 
    at $\sqrt{s}=2.76\,A$\,TeV.
    \label{F2}
    }
\end{figure}
%

As shown in Fig.~\ref{F2}, the event-averaged ellipticity $\epsilon_2$ and triangularity $\epsilon_3$ at the beginning of the hydrodynamic stage decrease as the length of the free-streaming period increases. However, for $\tau_s  >$ 6 fm/$c$, they increase again. Figure~\ref{F1} illustrates why this happens: around 5 fm/$c$ the free-streaming fireball begins to disintegrate and eventually separate into multiple pieces. This disintegration happens due to the absence of interactions between the partons which would otherwise keep the fluid together. So the matter distribution becomes more eccentric for large switching times, and the eccentricity coefficients $\epsilon_n$ increase accordingly.  

Free-streaming also drives the system out of equilibrium. For free-streaming (which corresponds to an infinite mean free path $\lambda_\mathrm{mfp}=\infty$), the Knudsen number 
\begin{eqnarray}
\mathrm{Kn} = \frac{\lambda_\mathrm{mfp}}{L_\mathrm{macro}},
\label{KnudsenNumber}
\end{eqnarray}
where $L_{macro}$ is the characteristic macroscopic length scale of the system, is infinite. This tells us that, even if the initial momentum distribution were thermal, the system would evolve further and further away from local thermal equilibrium. The inverse Reynolds number uses the hydrodynamic decomposition (\ref{TmunuHydroForm}) to describe how far away a system is from local thermal equilibrium. For a non-equilibrium pressure caused by shear viscosity, it is defined as the ratio between the scalar $\sqrt{\pi^{\mu\nu}\pi_{\mu\nu}}$ characterizing the magnitude of the shear stress and the thermal pressure $\mathcal{P}$:
\begin{eqnarray}
R_\pi^{-1} = \frac{\sqrt{\pi^{\mu\nu}\pi_{\mu\nu}}}{\mathcal{P}}.
\label{inverseReynoldsNum}
\end{eqnarray}

For an analytic estimate Eq.~(\ref{inverseReynoldsNum}) is inconvenient since the thermal pressure $\mathcal{P}$ is related to the energy density $e$ (whose initial profile can be calculated analytically in terms of the distribution function $f$) only through numerical lattice QCD calculations. This complication can be avoided by using a slightly different definition for the inverse Reynolds number:
\begin{eqnarray}
R^{-1} = \frac{\sqrt{\pi^{\mu\nu}\pi_{\mu\nu}}}{-\Delta^{\mu\nu}T_{\mu\nu}/3}
=\frac{\sqrt{\pi^{\mu\nu}\pi_{\mu\nu}}}{\mathcal{P}+\Pi}.\label{inverseReynoldsNum_new}
\end{eqnarray}
For a conformal EOS, $\mathcal{P}=\frac{1}{3}e$ and $\Pi=0$, this definition agrees with Eq.~(\ref{inverseReynoldsNum}), but if $\Pi\neq 0$ it allows for the following analytic computation of its initial value.
 
\begin{figure*}[!hbt]
    \includegraphics[width=0.9\linewidth]{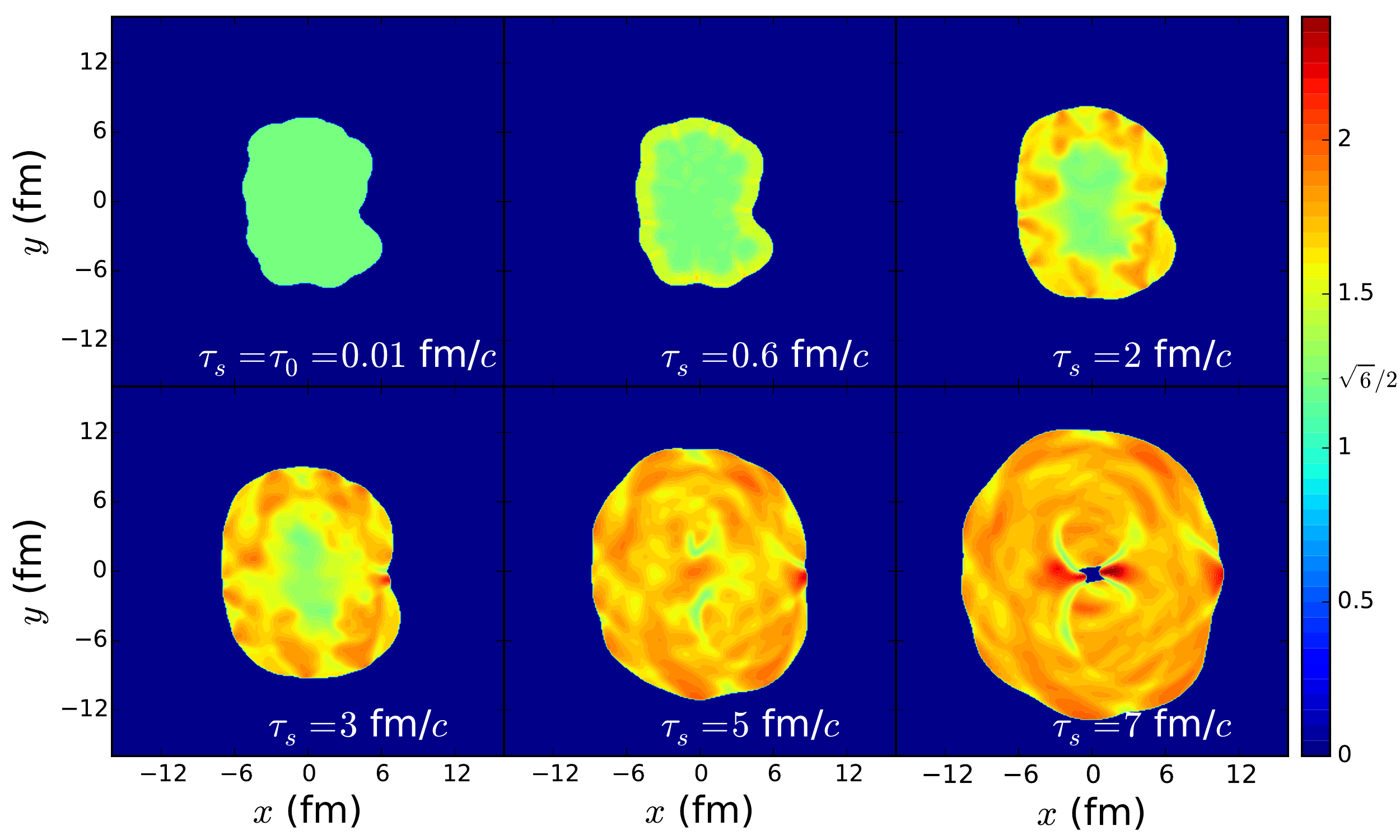}
    \caption{(Color online)
    Contour plots for the inverse Reynolds number $R^{-1}$ (\ref{inverseReynoldsNum_new})
    in the transverse plane just after the system is switched from free-streaming to 
    hydrodynamics, for several choice of switching time $\tau_s$, for a single fluctuating 
    Pb+Pb collision event of 10\% -- 20\% centrality at $\sqrt{s}=2.76$ $A$TeV. See text for 
    discussion.
    \label{F3}
    }
\end{figure*}

$R^{-1}$ is a local parameter whose initial value for the case at $\tau_s\eq\tau_0$ can be calculated analytically, provided the initial momentum distribution is locally isotropic in the transverse plane and the system is boost-invariant. In this case, inserting the definition (\ref{TmunuDef}) into (\ref{pimnExtraction}) to obtain
\begin{eqnarray}
&&\pi^{\mu\nu}\pi_{\mu\nu}\eq\!\int\!\frac{g\,d^3 p}{(2\pi)^3p^0} \!\int\!
     \frac{g\,d^3 p'}{(2\pi)^3p'^0} \;\Delta^{\rho\sigma}_{\alpha\beta}p^\alpha p^\beta 
     p'_\rho p'_\sigma f(p) f(p')
\nonumber \\ 
&&\eq\!\int\!\frac{g\,d^3 p}{(2\pi)^3p^0} \!\int\!\frac{g\,d^3 p}{(2\pi)^3p'^0} 
\l[({\bm{p}}{\cdot}{\bm{p}}')^2{-}\frac{1}{3}{\bm{p}}^2 {\bm{p}'}^2\r] f(p) f(p'),
\nonumber\\
\label{pimunuContraction}
\end{eqnarray}
using local transverse momentum isotropy to recast Eq.\,(\ref{boostInvariantF}) for massless partons in the transverse plane at $\eta_s=0$ into
\begin{eqnarray}
f(p)\big|_{\tau_0,\eta_s=0} 
 = \frac{\delta(y)}{\tau_0 \scaperp{p}}
\tilde{f}(\scaperp{p}),
\label{boostInvariantF_isotrpic}
\end{eqnarray}
rewriting $\int\frac{d^3 p}{p^0}=\int dy \,d^2\scaperp{p}$, and using the identity $p^0\eq\scaperp{p}$ for massless on-shell gluons we find
\begin{eqnarray}
&&\pi^{\mu\nu}\pi_{\mu\nu} \big|_{\tau_0,\eta_s{=}0} = g^2\!\int\frac{dy\, d^2p_\perp 
dy' d^2p'_\perp}{(2\pi)^6 \tau_0^2 p_\perp p'_\perp} 
\nonumber\\
&&\hspace*{2cm}
    \times \l[({\bm{p}}{\cdot}{\bm{p}}')^2{-}\frac{1}{3}{\bm{p}}^2 {\bm{p}'}^2\r] 
    \tilde{f}(p_\perp) \tilde{f}(p'_\perp) \delta(y) \delta(y') 
\nonumber \\
&&=\frac{g^2}{\tau_0^2}\int\frac{dp_\perp dp'_\perp}{(2\pi)^6} p_\perp^2 {p'}_{\!\!\perp}^2 
       \tilde{f}(p_\perp) \tilde{f}(p'_\perp)
\nonumber\\
&&\ \ \times
       \int \frac{d\phi_p d\phi_p'}{\text{sin}^2\phi_p \text{sin}^2\phi'_p}
       \bigl[ (\sin\theta_p\cos\phi_p\sin\theta'_p\cos\phi'_p\bigr.
\\\nonumber
&&\quad\bigl.{+}\sin\theta_p\sin\phi_p\sin\theta'_p\sin\phi'_p + \cos\theta_p\cos\theta'_p )^2 
     - \textstyle{\frac{1}{3}}\bigr]_{y=y'=0}. 
\label{pimunuContractionCalculation}
\end{eqnarray}
In the second line, $\bm{p}$ was decomposed as  ${\bm{p}} = p(\sin\theta_p \cos\phi_p, \sin\theta_p \sin\phi_p, \cos\theta_p)$ using spherical coordinates, with the polar angle $\theta_p$ ($\theta_p'$) being related to the pseudorapidity $\eta$ ($\eta'$) (which for massless partons agrees with their rapidity $y$ ($y'$)) through 
\begin{equation}
\cos\theta_p = \tanh\eta, \quad
\sin\theta_p = 1/\cosh\eta,
\label{angleTopseudoRapidity}
\end{equation}
since $\eta\eq{-}\ln\l[\text{tan}\l(\theta_p/2\r)\r]$. We see that for massless particles the angular integral can be factored from the integration over $\scaperp{p}$. At $y\eq{y'}\eq0$, $\theta_p\eq\theta'_p\eq\frac{\pi}{2}$, and the integration over azimuthal angles $\phi_p$ and $\phi_p'$ is easily performed, giving the result $2\pi^2/3$. Thus
\begin{eqnarray}
\pi^{\mu\nu}\pi_{\mu\nu}\big|_{\tau_0} = \frac{2\pi^2}{3} C^2
\label{largenessAtReynoldsNumber}
\end{eqnarray}
where $C\equiv \frac{g}{\tau_0}\int\frac{1}{(2\pi)^3} p_\perp^2 d\scaperp{p} \tilde{f}(\scaperp{p})$.

The value of $\mathcal{P}{+}\Pi$ (the denominator of $R^{-1}$) at $\tau\eq\tau_0$ and $\eta_s\eq0$ can be found from Eq.~(\ref{bulkPi}) (all quantities evaluated at $\tau\eq\tau_0$ and $\eta_s\eq0$):
\begin{eqnarray}
\mathcal{P}+\Pi&=& -\frac{1}{3}\Delta^{\mu\nu}T_{\mu\nu} \nonumber \\
& = & -\frac{1}{3}\Delta^{\mu\nu} \int\frac{g\, d^3 p}{(2\pi)^3 p^0} \, p_\mu p_\nu \delta(y)\frac{\tilde{f}(p)}{\tau_0\,\scaperp{p}} \nonumber \\
&=& \frac{g}{3(2\pi)^3\tau_0}\int \frac{d^2\scaperp{p}}{\scaperp{p}}p_\perp^2\tilde{f}(\scaperp{p})
= \frac{2\pi}{3} C.
\label{pAtReynoldsNumber}
\end{eqnarray}
Combining Eqs.~(\ref{largenessAtReynoldsNumber}) and (\ref{pAtReynoldsNumber}), the inverse Reynolds number $R^{-1}$ (\ref{inverseReynoldsNum_new}) at the matter formation time $\tau_0$ is seen to be independent of position in the transverse plane and equal to
\begin{eqnarray}
R^{-1} \big|_{\tau_0,\eta_s{=}0} = \sqrt{\frac{3}{2}} \approx 1.225.
\label{inverseReynoldsNumTau0}
\end{eqnarray}
This non-zero value for the initial inverse Reynolds number is caused by the anisotropy in the initial pressure whose longitudinal component vanishes due to the assumed zero width of the initial rapidity distribution ${\sim\,}\delta(y-\eta_s)$. It is consistent with the value obtained in anisotropic hydrodynamics which gives for a conformal system \cite{Bazow:2013ifa}:
\begin{eqnarray}
\frac{\left(\sqrt{\pi^{\mu\nu}\pi_{\mu\nu}}\right)_\mathrm{ahydro}}{\mathcal{P}}
=\sqrt{\frac{3}{2}}\frac{1-\mathcal{P}_L/\mathcal{P}_\perp}{1+\mathcal{P}_L/(2\mathcal{P}_\perp)},
\label{ahydro}
\end{eqnarray}
and thus agrees with our results at $\mathcal{P}_L=0$.\\

The value of $R^{-1}$ at the beginning of the hydrodynamic stage increases if the matter is allowed to free-stream for a finite time $\tau_s$ before switching to hydrodynamics. This is caused by the appearance of additional anisotropies in the transverse plane. As the partons free-stream and the matter expands outward, the transverse momentum distibution becomes locally anisotropic, especially in the regions near the outer edge of the fireball where the momenta point predominatly outward. This breaks the factorization in Eq.~(\ref{pimunuContractionCalculation}) of the angular integral from the one over the magnitude of $p_\perp$, and the transverse profile of $R^{-1}$ can no longer be calculated analytically. Fig.~\ref{F3} shows how the inverse Reynolds number $R^{-1}$ just after switching varies with $\tau_s$. The constant value of $R^{-1}$ shown in the first panel of this figure (where we switch directly at $\tau_0$) reflects the result (\ref{inverseReynoldsNumTau0}). As $\tau_s$ increases in the following panels, $R^{-1}$ first becomes large near the edge, but later also in the core of the profile, indicating that the entrie system is moving farther away from local thermal equilibrium if it is allowed to free-stream longer. 

One particular feature of the hydrodynamic initial conditions obtained by Landau matching after free-streaming is rather strong initial flow. Its radial part can be quantified by the mean radial velocity
\begin{eqnarray}
\{v_\perp\}=\frac{\int d^2r_\perp \gamma(\vecperp{r})\, v_\perp(\vecperp{r})\, e(\vecperp{r})}
                           {\int d^2r_\perp \gamma(\vecperp{r})\, e(\vecperp{r})},
\label{RadialVelocity}
\end{eqnarray}
where the curly bracket stands for the single-event average over the transverse plane. The $e$ and $v_\perp$ profiles are taken from the Landau matching results. For event-by-event hydrodynamical runs, the event-averaged initial radial velocity $\langle\scaperp{v}\rangle$ is obtained by averaging the value (\ref{RadialVelocity}) just after Landau matching over the events:
\begin{eqnarray}
\langle v_\perp \rangle = \frac{1}{N_{events}}\sum_{i=1}^{N_{events}}\{v_\perp\}^{(i)}.
\label{avgRadialVelocity}
\end{eqnarray}
%

%
\begin{figure}[!hbt]
    \includegraphics[width=0.9\linewidth]{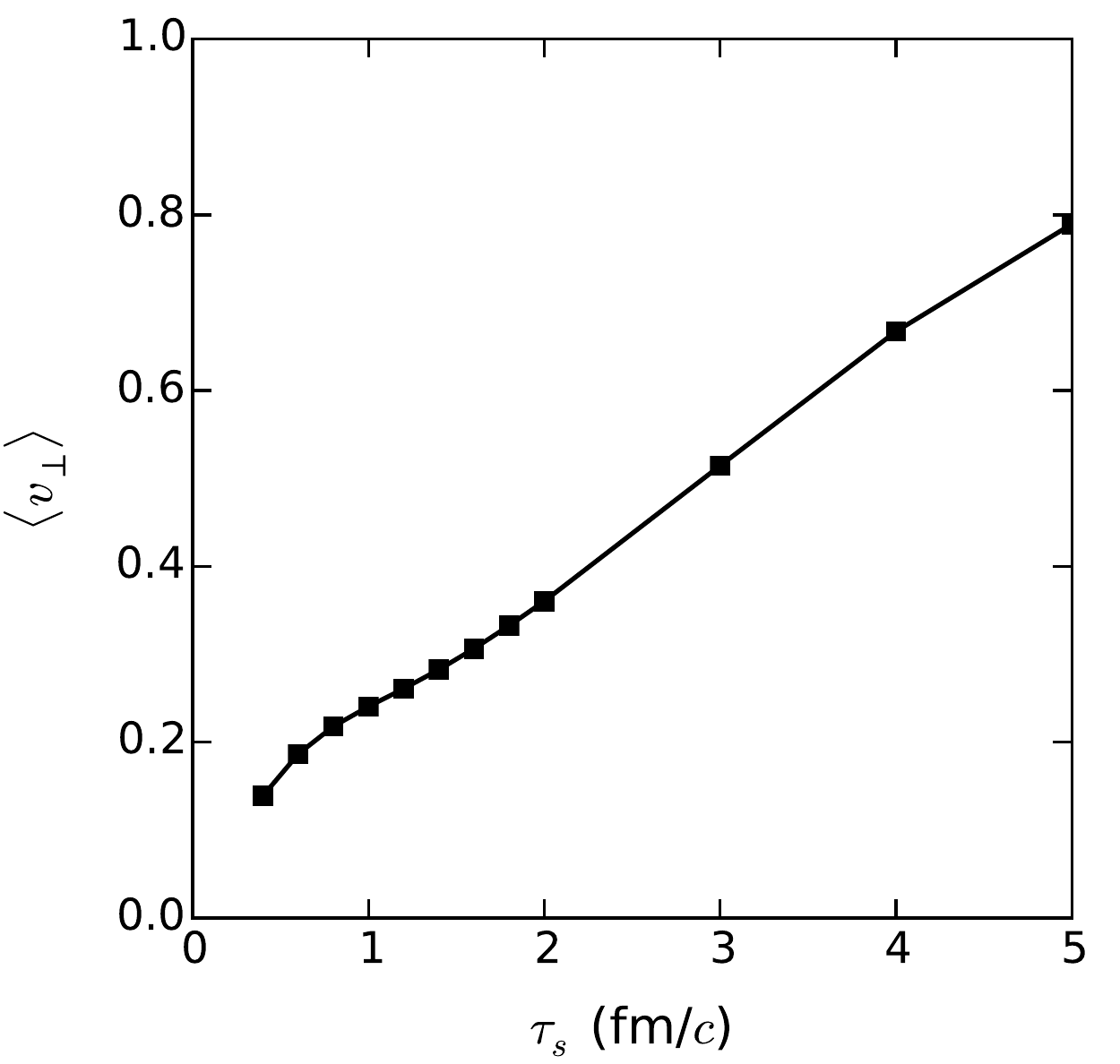}
    \caption{
    Event-averaged initial mean radial velocity as a function of switching time $\tau_s$, 
    for an ensemble of 400 fluctuating Pb+Pb collision events of 10\% -- 20\% centrality 
    at $\sqrt{s}=2.76\,A$\,TeV.
    \label{F4}
    }
\end{figure}
%

Fig.~\ref{F4} shows this quantity as a function of switching time. It initially rises very quickly, reaching 25\% of the speed of light already after 1 fm/$c$, and continues to grow at an approximate rate $\langle a_\perp \rangle \approx \frac{d \langle v_\perp \rangle }{d \tau_s} = 0.13$ $c^2$/fm over the next 5 fm/$c$. We reiterate that we expect a free-streaming pre-hydrodynamic stage to yield the largest possible pre-equilibrium effects on the subsequent hydrodynamic evolution. These will be discussed in the following sections; some of the trends we will observe in these sections may manifest themselves in weakened form when free-streaming will be replaced by more realistic pre-equilibrium dynamical models in future studies.

\section{Effects of free-streaming on hydrodynamical evolution}
\label{fsEffects}
\subsection{Radial flow}
\label{fsEffects:radialFlow}

Pre-equilibrium evolution endows hydrodynamics with significant initial flow but a reduced initial spatial eccentricity. The interplay between these two tendencies controls the radial flow and its anisotropies that are finally observed in the measured hadron momentum distributions.  For hydrodynamic simulations starting at different switching times, the initial conditions are rescaled to guarantee constant final total energy per unit space-time rapidity, $dE/d\eta_s$, as calculated from a single-shot hydrodynamic run with an event-averaged initial profile without pre-equilibrium evolution that has been tuned to reproduce the measured total charged hadron multiplicity density $dN_\mathrm{ch}/d\eta$ in central Pb+Pb collisions at $\sqrt{s}=2.76\,A$\,TeV. The final $dE/d\eta_s$ is calculated on the freeze-out surface as
\begin{equation}
\label{EqdEdetas}
\frac{dE}{d\eta_s} = \int_{\Sigma_\mathrm{fo}} T^{0\mu}(x) \,\frac{d^3\sigma_\mu(x)}{d\eta_s}.
\end{equation}
$dE/d\eta_s$ is roughly equal to the total multiplicity density $dN/d\eta$ multiplied by the mean 
transverse mass for charged hadrons, $\langle m_\perp \rangle^\mathrm{ch}$. Note that by varying initial parameters keeping $dE/d\eta_s$ fixed, we allow $dN_\mathrm{ch}/d\eta$ to vary: if the parameter change leads to an increase in radial flow, $\langle m_\perp \rangle^\mathrm{ch}$ increases and $dN_\mathrm{ch}/d\eta$ will decrease accordingly. The reason for using in this work the final energy $dE/d\eta_s$ rather than $dN_{ch}/d\eta$ for rescaling the initial distribution for each fluctuating event will be explained in Sec.~\ref{fsIssues}. 

\begin{figure}[h]
    \includegraphics[width=0.9\linewidth]{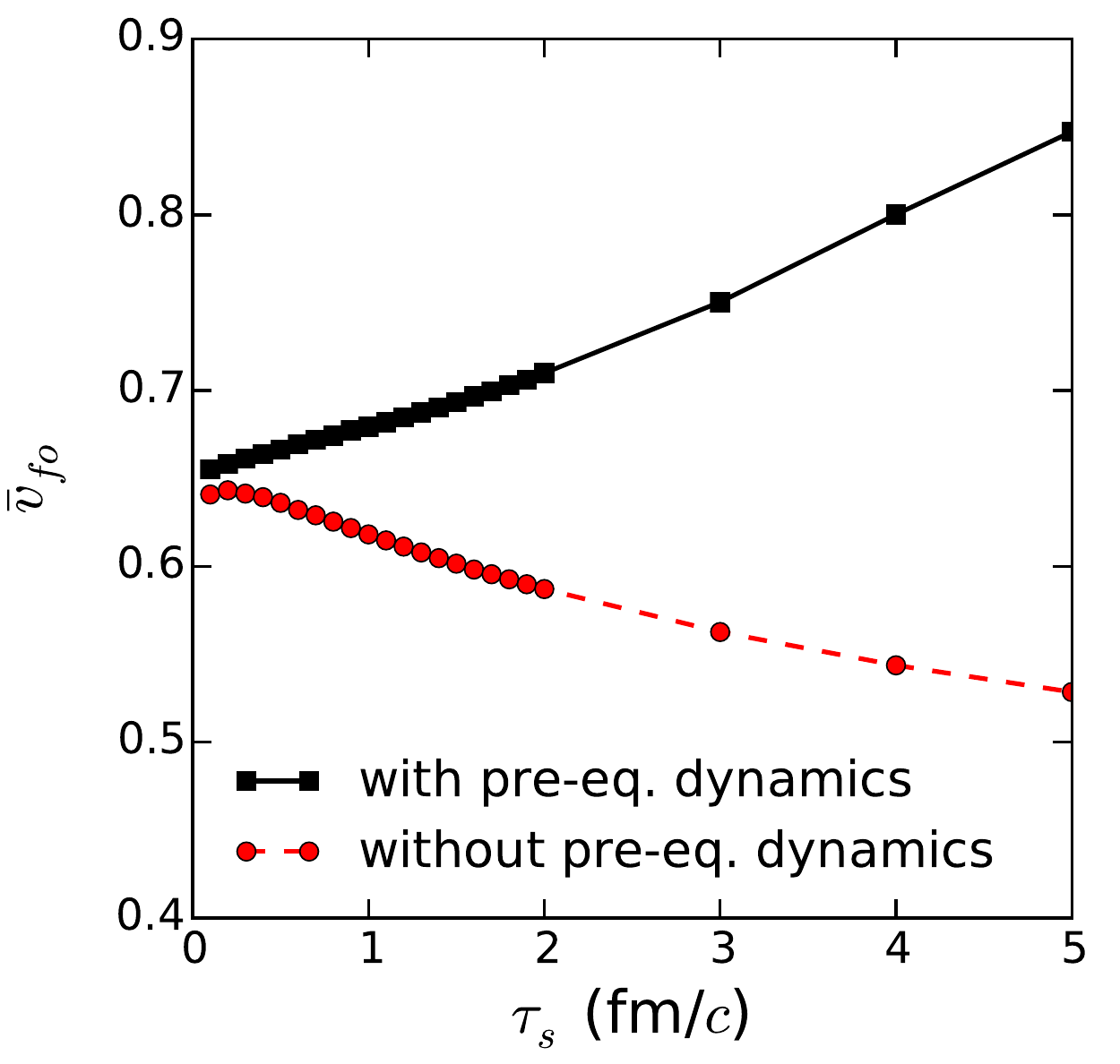}
\vspace*{-3mm}
    \caption{(Color online)
    $\bar{v}_{fo}$ as a function of switching time $\tau_s$, obtained from a single-shot 
    hydrodynamic event with and without pre-equilibrium evolution, for the same 
    ensemble-averaged MC-KLN initial profile. The solid (dashed) line corresponds 
    to including (excluding) pre-equilibrium flow before beginning of the hydrodynamic 
    evolution stage at $\tau_s$. 
    \label{F5}
    }
\end{figure}

We expect the increase with $\tau_s$ of the initial average radial flow shown in Fig.~\ref{F4} to manifest itself in a flow-induced blue shift of the finally measured hadron $p_\perp$-distributions, caused by the hydrodynamic radial flow on the freeze-out surface. For a single event, the final hydrodynamic radial flow can be characterized by the average radial velocity of the fluid cells on the freeze-out surface
\begin{eqnarray}
v_\mathrm{fo}\equiv\frac{\int_{\Sigma_\mathrm{fo}}u^\mu d^3\sigma_\mu\, v_\perp\, e}
                           {\int_{\Sigma_\mathrm{fo}}u^\mu d^3\sigma_\mu\, e} 
        = \frac{\int_{\Sigma_\mathrm{fo}}u^\mu d^3\sigma_\mu\, v_\perp}
                   {\int_{\Sigma_\mathrm{fo}}u^\mu d^3\sigma_\mu}.
\label{v_fo}
\end{eqnarray}
In the second equality we used that our freeze-out surface has constant temperature $T_\mathrm{dec}\eq120$\,MeV and, therefore, constant energy density $e_\mathrm{dec}$. The value $\bar{v}_{fo}$ of this quantity obtained from Eq.~(\ref{v_fo}) for a single hydrodynamic run with an {\emph{ensemble-averaged}} MC-KLN initial profile for Pb+Pb collisions at $\sqrt{s}\eq2.76\,A$\,TeV and 10\% -- 20\% centrality is shown by the black squares in Fig.~\ref{F5} as a function of switching time $\tau_s$. As expected from Fig.~\ref{F4}, it increases with $\tau_s$, contrary to what we observed earlier in simulations that ignored pre-equilibrium dynamics and started the hydrodynamic evolution with unevolved density profiles and zero transverse flow at the same $\tau_s$ (shown as red circles in Fig.~\ref{F5}).

%
\begin{figure*}[!hbt]
    \includegraphics[width=0.8\linewidth]{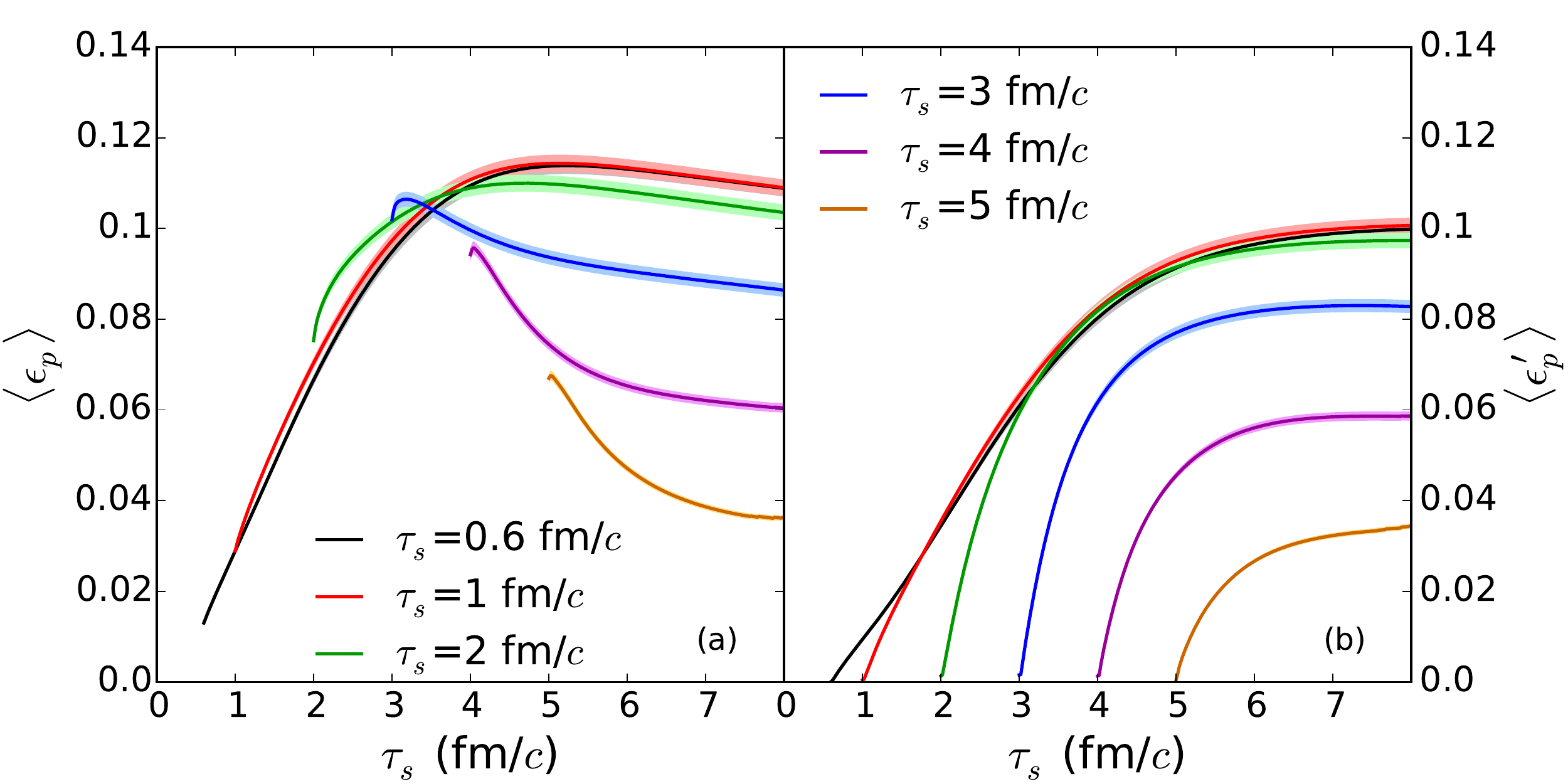}
    \caption{(Color online)
     Time evolution of the ensemble-averaged total hydrodynamic flow anisotropy 
     $\langle\epsilon_p\rangle$ (a) and total momentum anisotropy $\langle\epsilon'_p\rangle$
     (b) (see text for definitions) for different switching times, for an ensemble of 400 
     fluctuating Pb+Pb collision events of 10\% -- 20\% centrality at $\sqrt{s}=2.76$ $A$TeV.
     \label{F6}
    }
\end{figure*}
%

\subsection{Anisotropic flow}
\label{fsEffects:anisotropicFlow}

The total momentum anisotropy
\begin{eqnarray}
\epsilon_p'=\frac{\int d^2r_\perp (T^{xx}-T^{yy})}{\int d^2r_\perp (T^{xx}+T^{yy})},
\label{totalMomentumAnisotropy}
\end{eqnarray}
is directly and monotonously related to the elliptic flow $v_2^{\mathrm{ch}}$ of all charged hadron integrated over $\scaperp{p}$ \cite{Heinz:2005zg, Teaney:2009qa}. This quantity includes the collective flow anisotropy generated by pre-equilibrium dynamics as well as a contribution from the anisotropy of the local momentum distribution reflected in $\pi^{\mu\nu}$ \cite{Song:2007ux}. The collective flow part of this momentum anisotropy is captured by
\begin{eqnarray}
\epsilon_p=\frac{\int d^2r_\perp (T_\mathrm{id}^{xx}-T_\mathrm{id}^{yy})}
                          {\int d^2r_\perp (T_\mathrm{id}^{xx}+T_\mathrm{id}^{yy})},
\label{flowMomentumAnisotropy}
\end{eqnarray}
where $T^{\mu\nu}_\mathrm{id}=e u^\mu u^\nu{-}\mathcal{P}\Delta^{\mu\nu}$ is the ideal fluid part of the enery momentum tensor. Starting hydrodynamics at different switching times causes both quantities to saturate at different values. At large times, velocity shear effects die out and $\Pi,\,\pi^{\mu\nu}$ become small \cite{Song:2007ux}, and hence $\epsilon_p$ and $\epsilon'_p$ approach each other (see Fig.~\ref{F6}). Smaller saturated values of $\epsilon_p$ and $\epsilon'_p$ indicate weaker final anisotropic flow. Studying how the saturated values of these quantities change with switching time thus may help to constrain that parameter.

In order to calculate the ensemble-averaged total momentum anisotropy $\l<\epsilon_p'\r>$, we first rotate the transverse components of $T^{\mu\nu}$ in each event to maximize the magnitude of $\epsilon_p'$ and then sum over events. For $\l<\epsilon_p\r>$ we proceed similarly with the ideal fluid part of $T^{\mu\nu}$. Fig.~\ref{F6} shows how these ensemble-averaged $\langle \epsilon_p\rangle$ and $\langle \epsilon_p'\rangle$ evolve during the hydrodynamic stage, for several choices of the switching time. $\langle \epsilon_p'\rangle$ always starts out with zero magnitude at $\tau\eq\tau_s$ since the initial parton momentum distribution at $\tau_0$ is isotropic, and this isotropy of the spatially integrated momentum distribution is preserved by free-streaming. In contrast, $\epsilon_p$ starts out  after Landau matching with non-zero magnitude, as seen in Fig.~\ref{F6}a. This is due to anisotropies in the space-momentum correlations that were generated in the pre-equilibrium stage and that, after Landau matching, manifest themselves in anisotropies of the hydrodynamic flow profile $u^\mu(x)$. After the onset of hydrodynamic evolution, $\langle \epsilon_p'\rangle$ initially increases quickly, rapidly approaching $\epsilon_p$, and then saturates. For small switching times $\tau_s{\,<\,}$\,2 fm/$c$, $\langle \epsilon_p'\rangle$ has enough time to fully develop before freeze-out, and the saturated values show little sensitivity to $\tau_s$. However, for larger switching times $\tau_s>$ 2 fm/$c$, $\langle \epsilon_p'\rangle$ saturates at lower values that decrease rapidly with increasing $\tau_s$. If thermalization happens very late, hydrodynamics is no longer able to reach the same degree of final momentum anisotropy as obtained for early thermalization. This agrees with earlier findings in \cite{Kolb:2000sd, Heinz:2004pj}.

\section{Issues of parton-hadron conversion}
\label{fsIssues}

\begin{figure}[b!]
    \includegraphics[width=\linewidth]{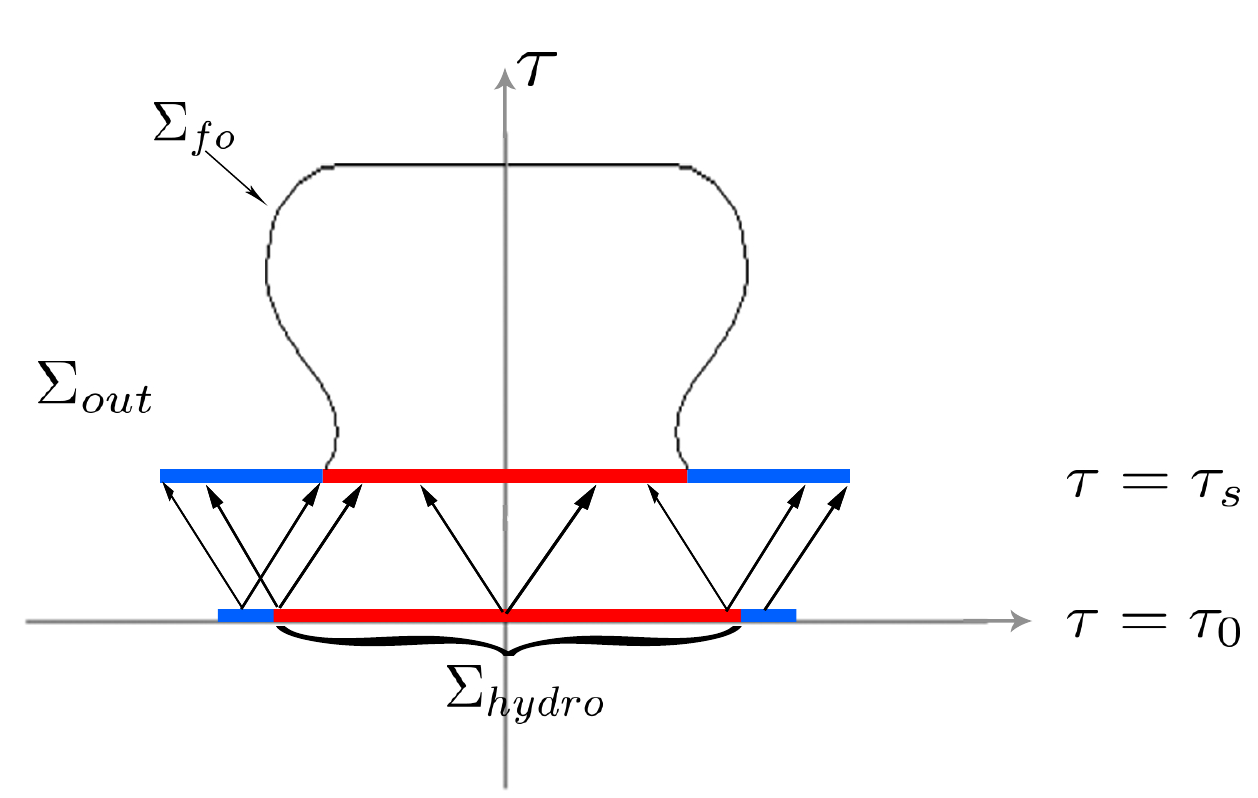}
    \caption{(Color online)
    Illustration of would-be fluid cells inside (red)  and outside (blue) of the freeze-out 
    surface during free-streaming in a (1+1)-dimensional space-time diagram where 
    the horizontal axis represents the transverse plane. 
    \label{F7}
    }
\end{figure}

As shown in Sec.~\ref{fsEffectsonICs}, free-streaming decreases the local energy density. As the free-streaming time $\tau_s$ increases, more cells fall with their energy density below the decoupling value $e_\mathrm{dec}$ even before the hydrodynamic stage starts. These partons will not be able to thermalize. The sketch shown in Fig.~\ref{F7} illustrates this phenomenon. In this figure, the thick colored horizontal lines represent areas occupied by partons. Blue segments stand for regions with energy density below $e_\mathrm{dec}$. After Landau matching, these regions would be outside the freeze-out surface $\Sigma_\mathrm{fo}$ that separates the thermalized fluid from free-streaming particles. The particles in those regions ``decouple'' instantaneously on a surface $\Sigma_\mathrm{out}$ that is part of the $\tau\eq\tau_s$ Landau-matching surface. Partons emitted from $\Sigma_\mathrm{out}$ neither thermalize nor evolve hydrodynamically. The red segment labeled as $\Sigma_\mathrm{hydro}$ indicates the region with $e>e_\mathrm{dec}$. After Landau-matching, cells on $\Sigma_\mathrm{hydro}$ become part of the thermalized fluid, so $\Sigma_\mathrm{hydro}$ forms the initial condition surface from which the hydrodynamic evolution starts. After hydrodynamic evolution, these cells decouple into free-streaming particles once they reach the freeze-out surface $\Sigma_\mathrm{fo}$, i.e. once their local energy density drops below $e_\mathrm{dec}$. 

At $\tau=\tau_0$, only a small region near the edge of the fireball is on $\Sigma_\mathrm{out}$. When free-streaming starts, the system expands both longitudinally and transversally, leading to a decrease of the local energy density, so $\Sigma_\mathrm{out}$ grows and $\Sigma_\mathrm{hydro}$ shrinks. At very large switching times $\tau_s{\,\gtrsim\,}8$\,fm/$c$, $\Sigma_\mathrm{out}$ covers the entire $\tau=\tau_s$ surface, and hydrodynamic evolution would never even start.

The non-thermalized partons on $\Sigma_\mathrm{out}$ hadronize directly from the free streaming stage. Contrary to the partons at $\Sigma_\mathrm{hydro}$ whose hadronization happens only after hydrodynamic evolution on $\Sigma_\mathrm{fo}$ and is described by the Cooper-Frye formula \cite{Cooper:1974mv}, partons on $\Sigma_\mathrm{out}$ do not have thermal distributions and we cannot use the Cooper-Frye formula to convert them to hadrons. Furthermore, most of the partons generated from the MC-KLN model are soft, with characteristic momenta ${<\,}1{-}2$\,GeV. No reliable models exist to convert such soft partons to hadrons. However, even though we cannot follow their evolution into final hadrons of well-defined mass and flavor, we can still follow their momentum and energy. So in the following sections, we propose to use the system's total energy flow distribution, instead of the momentum distributions of identified hadrons, to investigate radial and anisotropic flow for late switching times. We will trust our model's predictions for final identified hadron flows only when the contributions from $\Sigma_\mathrm{out}$ can be neglected.

Our inability to hadronize the contribution from $\Sigma_\mathrm{out}$ causes a problem for the rescaling of the initial profile. In common practice, the initial entropy profile is rescaled such that the calculated $dN_\mathrm{ch}/d\eta$ matches the experimental measurement. This procedure is problematic if their is a significant contribution of final hadrons emitted from $\Sigma_\mathrm{out}$ that cannot be included. Our way to work around this problem is to normalize the final {\em energy} on the transverse plane to a ``standard'' value. This ``standard'' final energy is obtained from a smooth fireball evolved with single-shot hydrodynamics starting at 0.6 fm/$c$ without pre-equilibrium dynamics, with parameters matched to reproduce the experimental $dN_\mathrm{ch}/d\eta$. Using longitudinal boost-invariance, i.e. the fact that the integrand can only depend on the rapidity difference $y{-}\eta_s$, the total energy per unit rapidity on the freeze-out surface is given by
\begin{eqnarray}
\left.\frac{dE}{dy}\right|_{\Sigma_\mathrm{fo}}&=&\sum_{i}\frac{g_i}{(2\pi)^3}
                          \int d^2p_\perp \int_{\Sigma_\mathrm{fo}} p^\mu d^3\sigma_\mu (n^\nu p_\nu) f_i
\nonumber\\
&=&\sum_{i}\frac{g_i}{(2\pi)^3}
                          \int dy\,d^2p_\perp \int_{\Sigma_\mathrm{fo}} p^\mu 
                          \frac{d^3\sigma_\mu}{d\eta_s} (n^\nu p_\nu) f_i
\nonumber\\
&=& \int_{\Sigma_\mathrm{fo}} n_\nu T^{\nu\mu} \frac{d^3\sigma_\mu}{d\eta_s} =
\left.\frac{dE}{d\eta_s}\right|_{\Sigma_\mathrm{out}},
\label{finalHydroEnergyiS}
\end{eqnarray}
where $n^\nu=(1,0,0,0)$ is the temporal unit vector in the lab frame and the sum runs over all species. The parton distribution for hadron species $i$ at the freeze-out surface can be written as
\begin{eqnarray}
f_i = f_{0,i} + \delta f_{\mathrm{shear}, i} + \delta f_{\mathrm{bulk}, i},
\label{f_decompose}
\end{eqnarray}
where $f_{0,i}$ is the local equilibrium distribution for hadron species $i$ while $\delta f_{\mathrm{shear}, i}$ and $\delta f_{\mathrm{bulk}, i}$) are the shear  and bulk viscous corrections
to $f_{0,i}$, accounting for the system's deviation from local thermal equilibrium. For the shear correction we use the ansatz \cite{Teaney:2003kp, Baier:2006um}
\begin{eqnarray}
\delta f_{\mathrm{shear}, i} = f_{0,i} (1{\pm}f_{0,i})
\frac{\pi^{\mu\nu} p_{\mu} p_{\nu}}{2T^2(e+\mathcal{P})}.
\label{f_shear}
\end{eqnarray}
For the bulk correction we use the following expression, derived from the 14-moment approximation  for particles with Boltzmann statistics \cite{Monnai:2009ad}:
\begin{eqnarray}
\delta f_{\mathrm{bulk}, i} = -f_{0,i}\Pi\l[B_{0,i}\,m_i^2+D_{0,i}\,u^\mu p_{\mu}+E_{0,i}\,(u^\mu p_{\mu})^2\r],\nonumber\\
\label{f_bulk}
\end{eqnarray}
For a non-interacting hadron resonance gas the coefficients $B_0(T)$, $D_0(T)$ and $E_0(T)$ were calculated in Ref.~\cite{Noronha-Hostler:2013gga} in the Boltzmann limit. 

Although in our work here the bulk viscous pressure approaches zero on the short bulk relaxation time scale $\tau_\Pi$, its initial value from the Landau matching is large, and the bulk viscous correction $\delta f_{\mathrm{bulk}, i}$ remains significant over the early part of the hydrodynamic freeze-out surface sketched in Fig.~\ref{F7}. To match the last two lines of Eq.~(\ref{finalHydroEnergyiS}) it is therefore important to include $\delta f_{\mathrm{bulk}, i}$ in the definition of the distribution function $f_i$.%
\footnote{%
     {For a quantitatively accurate matching one actually must correct the ansatz 
     (\ref{f_bulk}) and the calculation \cite{Noronha-Hostler:2013gga} of its 
     coefficients $B_0(T)$, $D_0(T)$ and $E_0(T)$ for quantum statistical effects. 
     We found a 9\% discrepancy between the value of $\Pi$ obtained by 
     reconstructing it from $\delta f_{\mathrm{bulk}, i}$ using its kinetic definition 
     \cite{Monnai:2009ad} and the value obtained directly by applying the projection 
     (\ref{bulkPi}) on the hydrodynamic energy-momentum tensor. Since in our case
     effects from $\Pi$ on the hadron spectra and flow coefficients, once integrated over
     the entire freeze-out surface, are small, we here ignored this 9\% discrepancy.
     }}

Even if we do not know how to hadronize the partons on $\Sigma_\mathrm{out}$, we can easily add their contribution to $dE/dy$. At $y=\eta_s=0$, we find
\begin{eqnarray}
&&\left.\frac{dE}{dy}\right|_{\Sigma_\mathrm{out}} =\frac{g}{(2\pi)^3}\int d^2\scaperp{p} \int_{\Sigma_\mathrm{out}}p^\mu d^3\sigma_{\mu} (u^\nu p_\nu) f(x,p) 
\nonumber \\
&&=\int_{\Sigma_\mathrm{out}} n_\nu T^{\nu\mu}\frac{d^3\sigma_\mu}{d\eta_s} 
            = \tau_s\int_{\Sigma_\mathrm{out}} d^2r_\perp T^{00} =\left.\frac{dE}{d\eta_s}\right|_{\Sigma_\mathrm{out}}\!\!\!\!\!.
\label{finaloutEnergy}
\end{eqnarray}
Here $f(x, p)$ is the distribution function for the initial free-streaming partons, and we again used the property $f(x, p) \sim \delta(y-\eta_s)$ to convert
\begin{eqnarray}
d^2p_\perp d^3 \sigma_\mu
&=& (dy\, d^2p_\perp) \frac{d^3\sigma_\mu}{d\eta_s} 
= n_\mu (dy\, d^2p_\perp) (\tau_s d^2r_\perp).\qquad
\end{eqnarray}
The final total energy after free-streaming and hydrodynamical evolution is the sum of the contributions from $\Sigma_\mathrm{out}$ and the hydrodynamic freeze-out surface $\Sigma_\mathrm{fo}$. We can now rescale the initial gluon distribution function such that for each switching time $\tau_s$ the final total energy reproduces the ``standard'' value defined above. We repeat that holding the final energy $dE/dy$ fixed is not equivalent to demanding fixed final multiplicity $dN_{ch}/d\eta$ (see discussion in Sec.~\ref{fsEffects:radialFlow}). This should be kept in mind when interpreting the results from the following study of the sensitivity of physical observables to the switching time $\tau_s$ (i.e. to the time the system needs to thermalize sufficiently for hydrodynamics to become applicable).

\begin{figure}[tb]
    \includegraphics[width=0.9\linewidth]{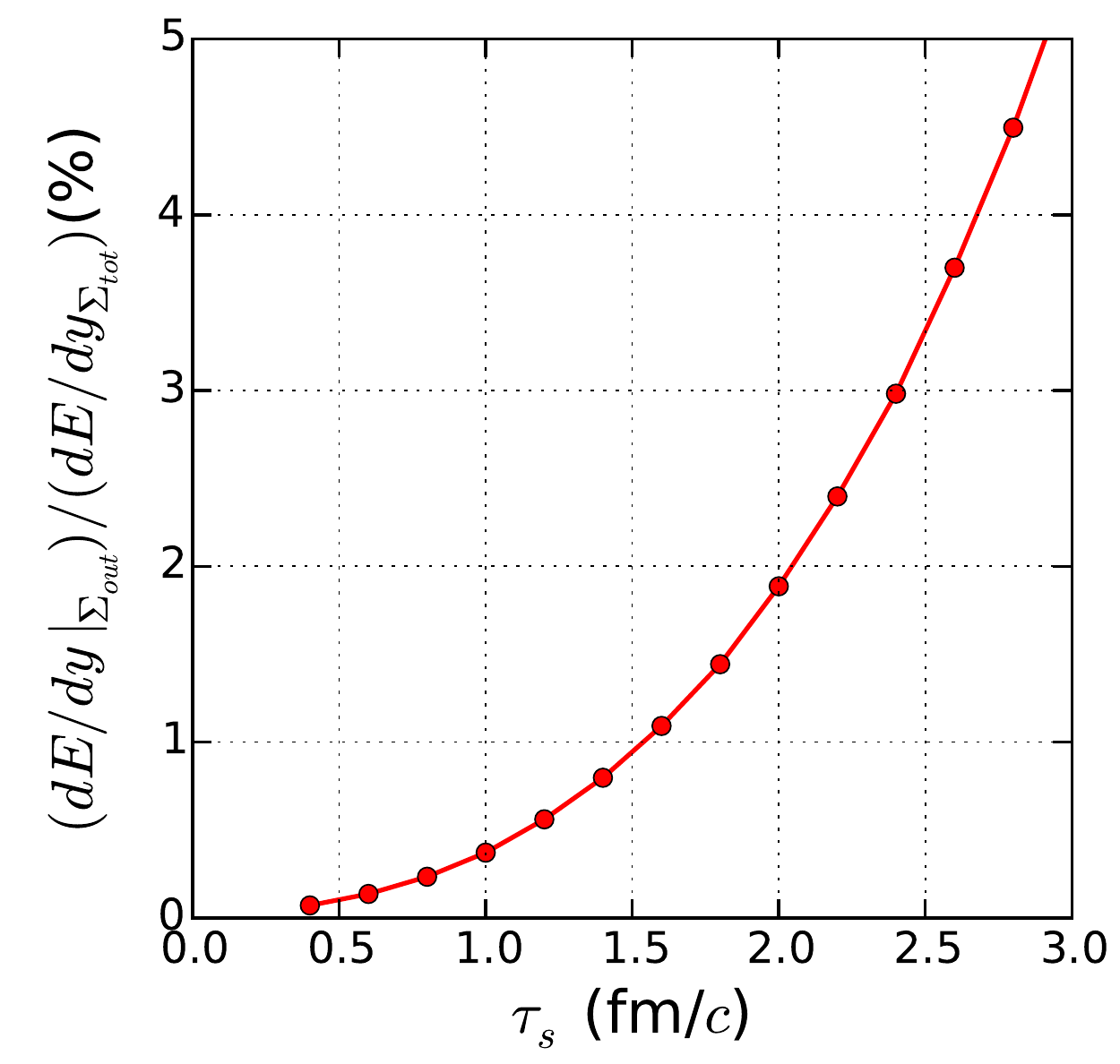}
    \caption{(Color online)
    The fraction (in percent) of the total final energy $\frac{dE}{dy}$ contributed by 
    non-thermalized particles on $\Sigma_{out}$, as a function of the switching time $\tau_s$.
    \label{F8}
    }
\end{figure}

Because of our inability to convert the partons on $\Sigma_{out}$ to hadrons, we do not know how to include their contribution in final hadronic observables such as transverse momentum spectra and flow anisotropies. These observables are computed by only including hadrons emitted from $\Sigma_{fo}$:
\begin{eqnarray}
\frac{dN_i}{dy d\phi_p} &=& \frac{g_i}{(2\pi)^3}\int p_\perp dp_\perp 
\int_{\Sigma_\mathrm{fo}} \!\!\! p^\mu d^3\sigma_\mu(x)\, f_i(x, p),\quad
\label{hadronSpectra} 
\\
v_n e^{i n \Psi_n} &=& \frac{\sum_i \int_{-\pi}^{\pi}d\phi_p \frac{dN_i}{dy d\phi_p} e^{i n \phi_p}}{\sum_i \int_{-\pi}^{\pi}d\phi_p \frac{dN_i}{dy d\phi_p}}\label{hadronicFlow},
\end{eqnarray}
where the sum over $i$ runs over all hadron species. Obviously, theoretical predictions based on this procedure are not trustworthy if the neglected contribution from $\Sigma_\mathrm{out}$ presents a significant fraction of the total energy. Fig.~\ref{F8} shows that the associated error increases rapidly with $\tau_s$, but remains below 5\% for $\tau_s{\,\lesssim\,}3$\,fm/$c$. If the system takes longer than 3 fm/$c$ to thermalize, we can no longer ignore the ``corona'' \cite{Werner:2007bf} of particles emerging from $\Sigma_\mathrm{out}$. We will therefore consider comparisons of our predictions for hadronic observables with experimental data ``meaningful'' only for runs with $\tau_s{\,\lesssim\,}3$\,fm/$c$.

\section{Energy flow anisotropy}
\label{energyFlow}

Due to the lack of the contribution from partons on $\Sigma_\mathrm{out}$,  for large switching times we are unable to fully reconstruct the hadron azimuthal distribution and its flow anisotropy. This is a big handicap when trying to explore the effect of large switching times on those physical observables: The largest effects one sees are due to the loss of particles through the $\Sigma_\mathrm{out}$ surface.

One way around this problem is to construct the flow anisotropy from the azimuthal distribution of energy instead of that of the hadrons. Since the energy angular distribution closely tracks that of the particles, the energy flow anisotropies should be strongly correlated with the (momentum-integrated) hadron flow anisotropies. In this section, we construct the energy flow anisotropy and calibrate it by comparing it to the hadron flow anisotropy for small switching times, when the contributions from $\Sigma_\mathrm{out}$ are negligible.

The azimuthal distributions of energy emitted from $\Sigma_\mathrm{out}$ and $\Sigma_\mathrm{fo}$ at $y\eq\eta_s\eq0$ are given by
\begin{eqnarray}
\left.\frac{dE}{dyd\phi_p}\right|_{\Sigma_\mathrm{fo}}\!\!\!
&=& \sum_{i}\frac{g_i}{(2\pi)^3}\int_{0}^{\infty} \!\!\! n^\nu p_\nu \scaperp{p} d\scaperp{p}\int_{\Sigma_\mathrm{fo}} \!\!\!\! p^\mu d^3\sigma_\mu f_i, \quad\ \ 
\label{dEdydphi_fo}\\
\left.\frac{dE}{dyd\phi_p}\right|_{\Sigma_\mathrm{out}}\!\!\!\!\!
&=& \frac{g\,\tau_s}{(2\pi)^3}\int_{0}^{\infty} \!\!\!n^\nu p_\nu p_\perp^2 dp_\perp
        \int_{\Sigma_\mathrm{out}} \!\!\!\! d^2r_\perp f. 
\label{dEdydphi_out}
\end{eqnarray}
In (\ref{dEdydphi_fo}) $i$ runs over all hadron species, while (\ref{dEdydphi_out}) contains only the gluons generated from the KLN model. In (\ref{dEdydphi_out}) we used that on $\Sigma_\mathrm{out}$, $p^\mu d^3\sigma_\mu\eq{p}_\perp \tau_s d^2r_\perp$ for massless partons at $y\eq0$. Note that at $y\eq0$, $n^\nu p_\nu\eq\scaperp{p}$ for the massless partons in (\ref{dEdydphi_out}) while $n^\nu p_\nu\eq{m}_{\perp, i}\eq\sqrt{m_i^2{+}p_\perp^2}$ for the hadron species $i$ in (\ref{dEdydphi_fo}).

\begin{figure}[ht]
    \includegraphics[width=\linewidth]{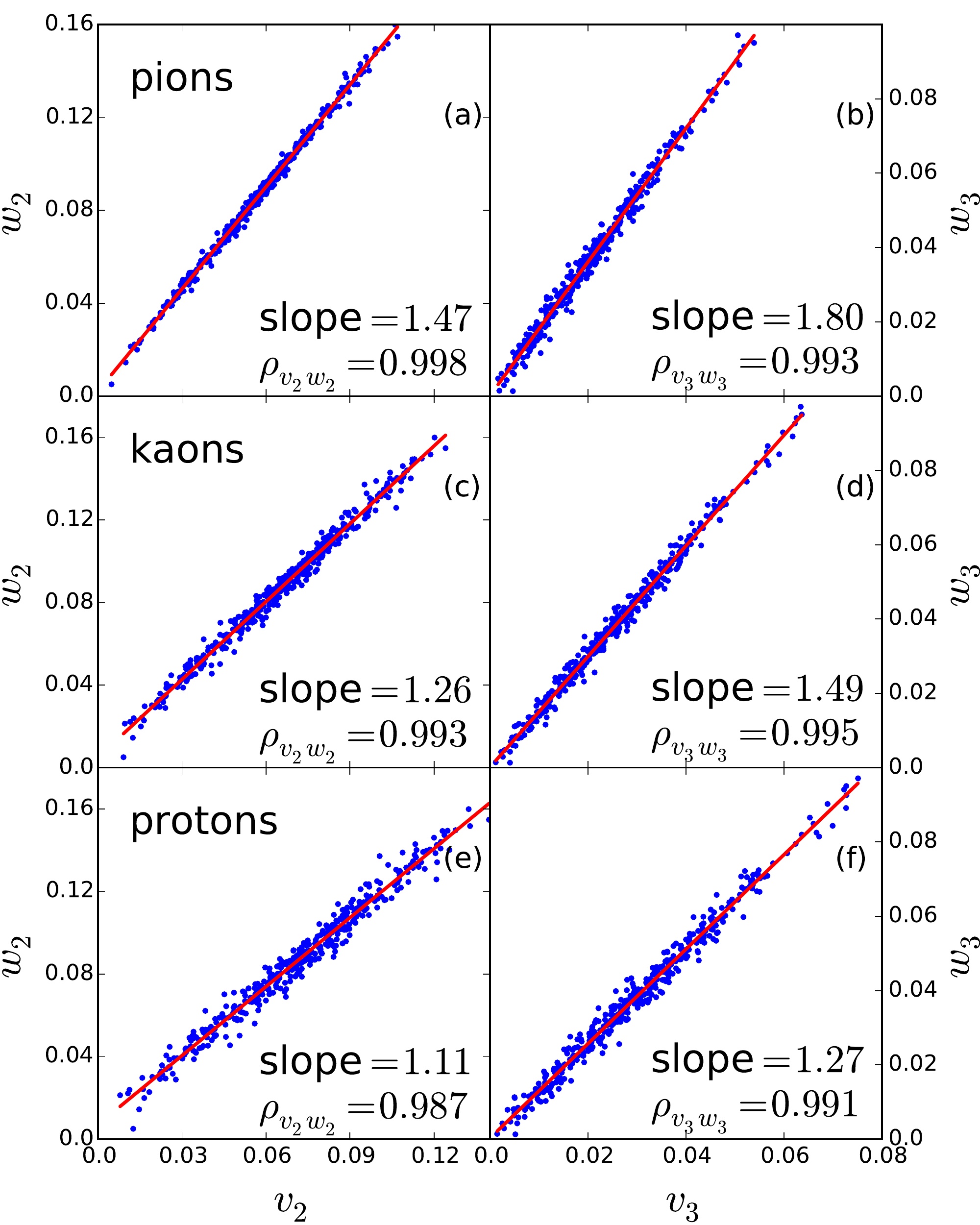}
    \caption{(Color online)
    Scatter plot of the hadron flow anisotropies $v_{2,3}$ for pions (top panels), kaons 
    (middle panels) and protons (bottom panels), and energy flow anisotropies $w_{2,3}$ 
    from 400 fluctuating hydrodynamic events using $\tau_s\eq0.6$\,fm/$c$ for the switching 
    time. The left (right) column is for elliptic (triangular) flow. The slope and correlation 
    coefficients from a linear fit to the scatter plots are noted in each panel.
    \label{F9}
  }
\end{figure}

Summing these two contributions to a single distribution $dE/dyd\phi$, the Fourier coefficients of this azimuthal energy distribution can be extracted by the same procedure as used for calculating hadronic anisotropic flow coefficients:
\begin{eqnarray}
w_n e^{i n \bar{\Psi}_n} = \frac{\int \frac{dE}{dyd\phi_p} e^{i n \phi_p}\;d\phi_p}{\int \frac{dE}{dyd\phi_p} \;d\phi_p}.
\label{energyFlowAnisotropy}
\end{eqnarray}
$w_n$ is the energy flow anisotropy coefficient, which quantifies the azimuthal distribution of the energy contributed by all the particles -- non-thermalized partons as well as frozen-out, hydrodynamically flowing hadrons. $\bar{\Psi}_n$ is the energy flow angle associated with $w_n$.

\begin{figure*}
    \includegraphics[width=0.495\linewidth]{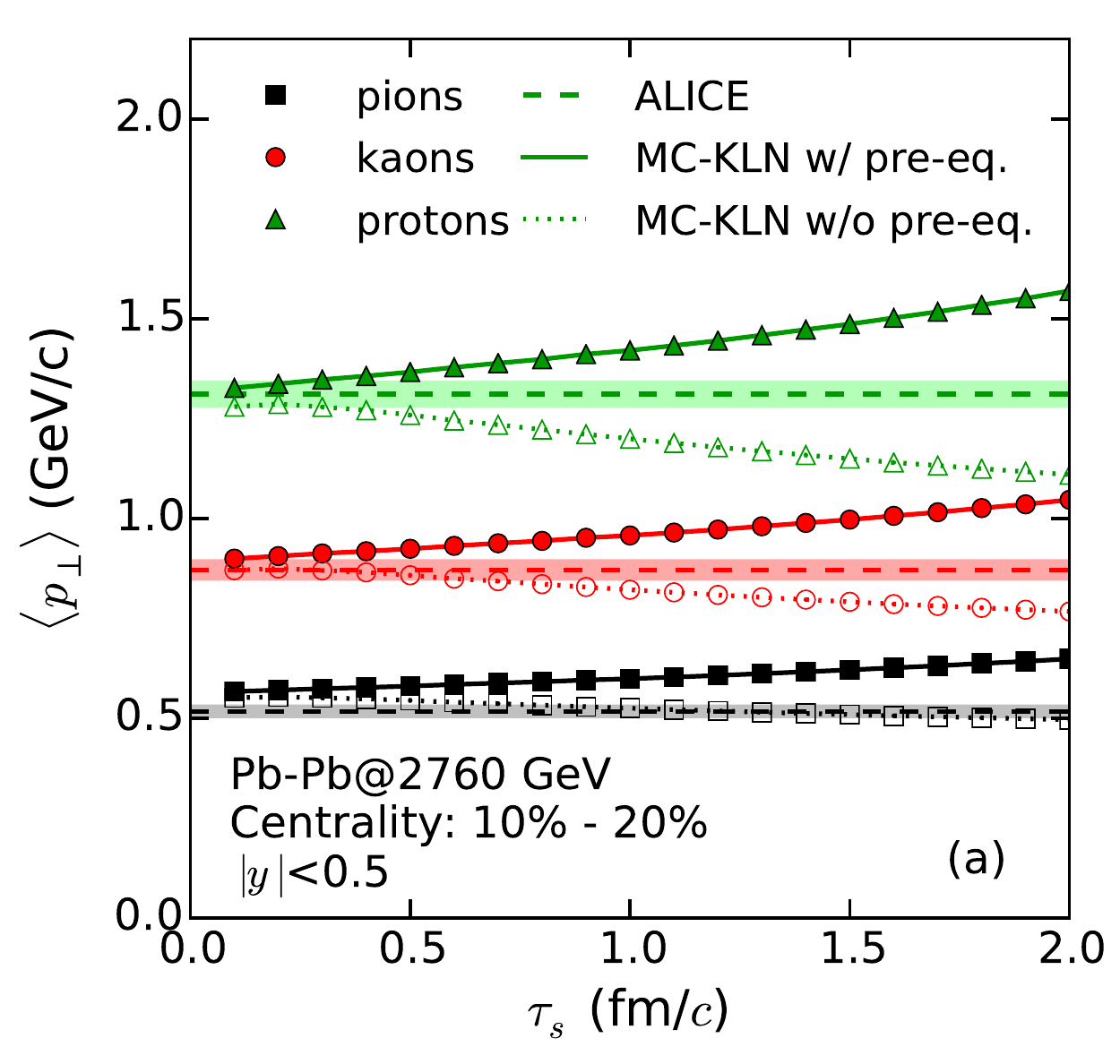}
    \includegraphics[width=0.495\linewidth]{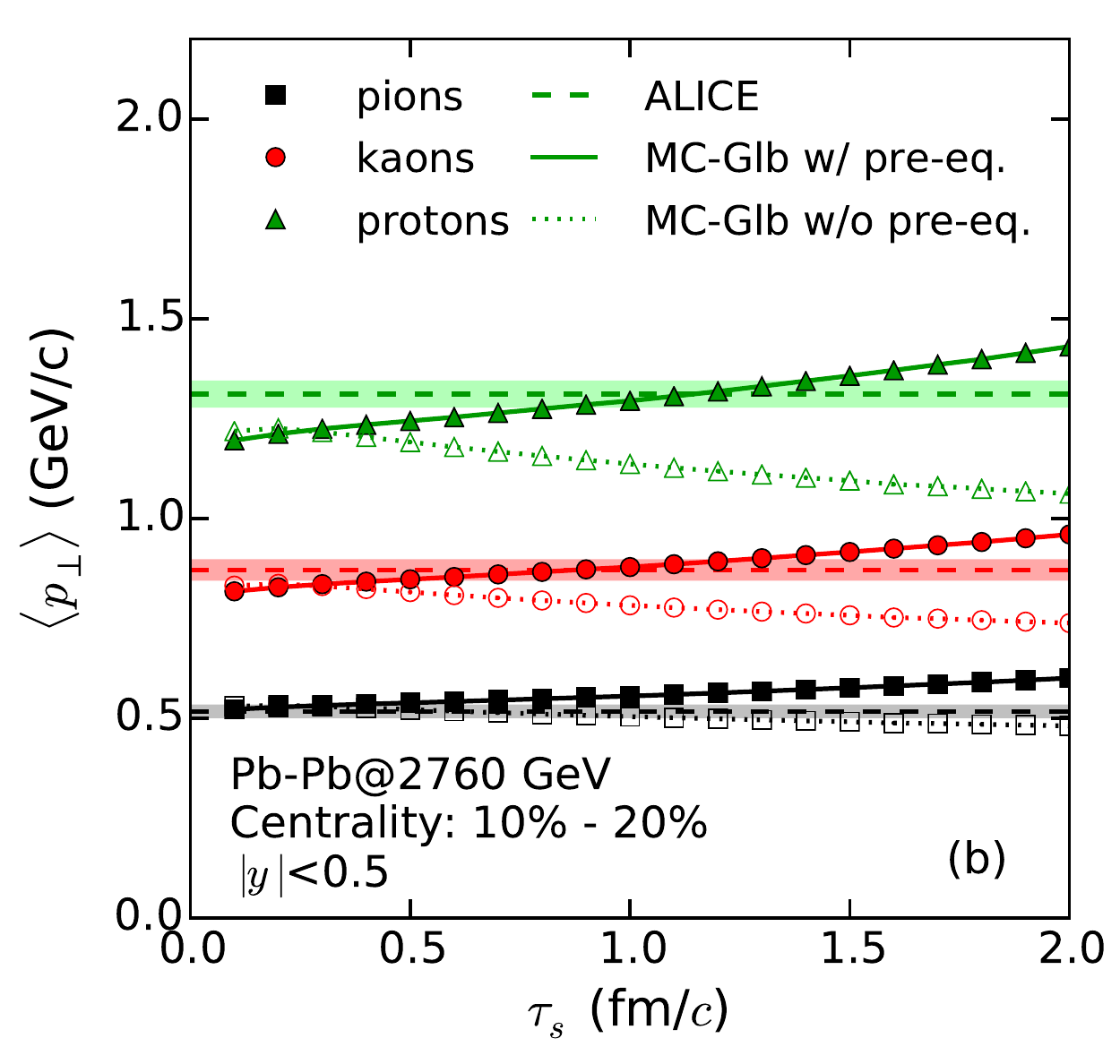}
    \caption{(Color online)
    Mean transverse momentum for pions, kaons and protons, with and without 
    pre-equilibrium dynamics, compared with data from the ALICE collaboration 
    \cite{Abelev:2013vea} (horizontal bands), for both MC-KLN (a) and MC-Glauber (b)
    initial models. 
    \label{F10}
    }
\end{figure*}

We will now show that, for small $\tau_s$ whose the contributions from $\Sigma_\mathrm{out}$ to energy and particle emission are negligible, $w_n$ from Eq.~(\ref{energyFlowAnisotropy}) and $v_n$ from Eq.~(\ref{hadronicFlow}) are tightly correlated. The advantage of $w_n$ over $v_n$ is that it is easy to include all contributions, including that from $\Sigma_\mathrm{out}$, in our calculation of the energy flow, while $v_n$ only accounts for contributions from $\Sigma_{fo}$ and thus misses a large fraction of the emitted hadrons when $\tau_s$ is large. If $w_n$ and $v_n$ are tightly correlated for small $\tau_s$, where both $v_n$ and $w_n$ account for essentially all emitted particles, we are allowed to use $w_n$ as a proxy for the true $v_n$ also for large $\tau_s$, where $v_n$ computed from (\ref{hadronicFlow}) no longer faithfully represents the full system. 

Fig.~\ref{F9} shows the correlation between the hadron flow anisotropies $v_n^i$ for three selected particle species $i$ and the energy flow anisotropies $w_n$ for harmonic orders $n\eq2$ and 3. For this comparison we chose $\tau_s\eq0.6$\,fm/$c$ to guarantee that the contribution to $w_n$ from $\Sigma_\mathrm{out}$ is negligible. We observe almost perfect correlations, with correlation coefficients very close to 1, for both $w_2$ vs. $v_2$ and $w_3$ vs. $v_3$. The slopes $w_n/v_n^i$ are larger than 1 for all hadron species $i$, demonstrating a stronger sensitivity of the energy flow coefficients $w_n$ to the hydrodynamic flow anisotropies than of $v_n$ for individual hadrons. Among the hadrons, heavier species such as protons are more sensitive to hydrodynamic flow than lighter species \cite{Kolb:2000sd}, but even for protons $v_{2,3}$ are still smaller than $w_{2,3}$. We confirmed similarly strong correlations between $v_{2,3}$ and $w_{2,3}$ at other switching times $\tau_s{\,<\,}2.5$\,fm/$c$ but saw that the correlation gradually breaks down for large $\tau_s$ values $\tau_s{\,>\,}3$\,fm/$c$ when too much of the total energy emerges from $\Sigma_\mathrm{out}$.

\section{Constraining the duration of the pre-equilibrium stage}
\label{upperLimitOfFS}

In this section, we will use the energy flow anisotropies $w_{2,3}$ and the mean transverse momenta for pions, kaons and protons to constrain the duration of the pre-equilibrium stage. It is well known that the average transverse momenta of hadrons with different masses help to separate random thermal motion (i.e. the freeze-out temperature) from the effect of collective radial hydrodynamic flow in the final state. Switching from initial free-streaming to hydrodynamics at later times means more initial flow after Landau matching but less time for developing hydrodynamic flow. Fig.~\ref{F5} has already shown that the net effect is an increased radial flow at freeze-out which will lead to harder momentum spectra and an increase in the average $p_\perp$. It is therefore expected that the measured $\langle p_\perp\rangle$ values for pions, kaons and protons will put an upper limit on the switching time, by limiting the amount of radial flow at freeze-out. 

This is illustrated in Fig.~\ref{F10} where the mean transverse momenta $\langle p_\perp \rangle$ for pions, kaons and protons are plotted as a function of switching time $\tau_s$, with MC-KLN initial conditions (propagated with specific shear viscosity $\eta/s\eq0.2$) in panel (a) and MC-Glauber initial conditions (propagated with $\eta/s\eq0.08$) in panel (b). Shown for comparison are experimental data from the ALICE Collaboration \cite{Abelev:2013vea} that were obtained by extrapolating the measured spectra to the full $p_\perp$ range before calculating the mean. When pre-equilibrium dynamics is included in the calculations (solid lines with filled symbols), the mean transverse momenta are seen to increase with $\tau_s$, as anticipated. The effect is strongest for protons whose large mass makes them most susceptible to flow. One sees that, with the chosen values for the shear viscosity and freeze-out temperatures, the MC-KLN initializations with free-streaming pre-equilibrium dynamics have difficulties accommodating the data unless one postulates essentially instantaneous thermalization, and even then the pion mean $\scaperp{p}$ is still statistically significantly too large (by about 10\%) compared to the measurements. On the other hand, the smaller shear viscosity used for evolving the MC-Glauber initial profiles in panel (b) reduces the transverse shear stress and thus builds less radial flow, giving room for some pre-equilibrium radial flow. With this combination of initial conditions and shear viscosity, a switching time around 1\,fm/$c$ appears to be preferred over significantly smaller and larger $\tau_s$ values.

In Fig.~\ref{F10} we also show for comparison as dashed lines with open symbols the corresponding results for hydrodynamic evolution without pre-equilibrium dynamics. In this case $\tau_s$ has the meaning of the starting time for the hydrodynamic evolution, but with initial conditions that have not evolved in the transverse plane between $\tau_0$ and $\tau_s$. Without pre-equilibrium, delaying the start of the hydrodynamic expansion leads to a reduction of the final radial flow, since less time is available for its generation before the matter reaches the decoupling temperature. As a result, $\langle p_\perp\rangle$ decreases with increasing $\tau_s$, and the effect is again stronger for protons than for pions and kaons, due to their larger mass. 

\begin{figure}[b!]
\begin{center}
    \includegraphics[width=0.9\linewidth]{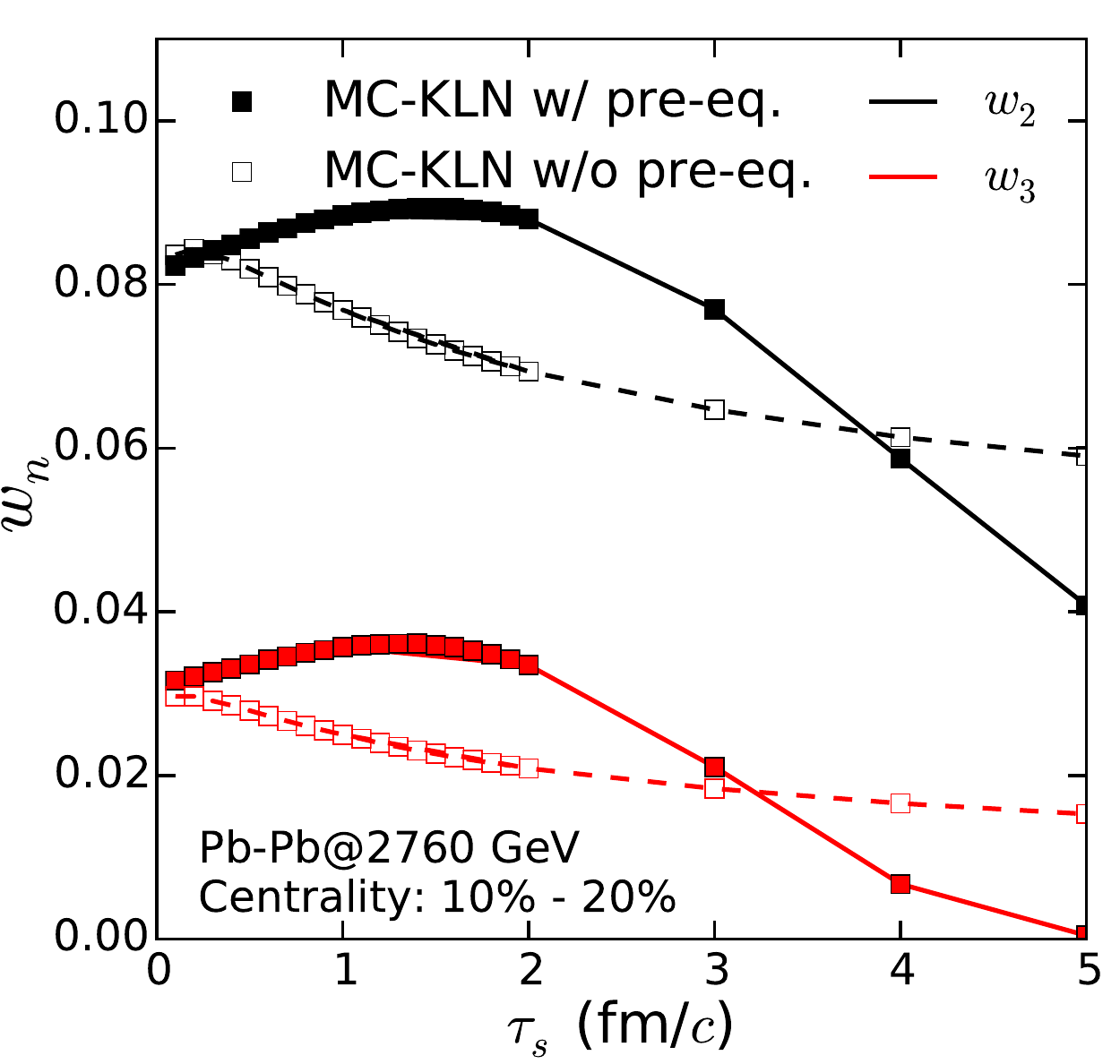}
\end{center}
    \caption{(Color online)
    Elliptic and triangular energy flow anisotropy coefficients 
    $w_2$, $w_3$ as a function of the switching 
    time from single-shot hydrodynamic simulations with a smooth ensemble-averaged MC-KLN initial profile.
    \label{F11}
    }
\end{figure}

For the purely hydrodynamic runs without pre-equilibrium dynamics, we see in Fig.~\ref{F10} that no choice of $\tau_s$ can reproduce all three measured $p_\perp$ values simultaneously, for either of the two initial condition models and associated shear viscosities. The reader may wonder how this can be consistent with successful earlier fits of the measured transverse momentum spectra for these three particle species \cite{Shen:2011eg,Abelev:2012wca}. Part of the answer is that in \cite{Shen:2011eg,Abelev:2012wca} the quality of the model description of the data was judged by the overall shape of the $p_\perp$-distributions whereas here we compare with only a single moment of that distribution which, however, measured with very good precision. We will return to this question in the following section where we look at a somewhat larger set of experimental data that are, in addition to the mean $p_\perp$, also sensitive to the shape of the $p_\perp$-distributions and try to fit them by simultaneously varying several hydrodynamic parameters.

For anisotropic flow the $\tau_s$ dependence is more subtle: as shown in Fig.~\ref{F2}, a free-streaming pre-equilibrium stage reduces the source eccentricity at the start of the hydrodynamic evolution, so it reduces the amount of flow anisotropy that can be generated during the hydrodynamic stage in response to this initial eccentricity. However, as shown in Fig.~\ref{F6}a, it also creates a non-zero hydrodynamic flow anisotropy at the start of the fluid stage which gives the hydrodynamic evolution of anisotropic flow a boost. Fig.~\ref{F11} shows that, for switching times up to about 2\,fm/$c$, the combined effect are final elliptic and triangular flow anisotropies $w_{2,3}$ that are almost independent of the switching time. In fact, both $w_2$ and $w_3$ slightly increase with increasing switching time until $\tau_s$ reaches about 1.5\,fm/$c$. Only for larger switching times beyond 2\,fm/$c$ does the reduction of $\varepsilon_{2,3}$ before the start of the hydrodynamic evolution cut into the finally established anisotropic flow coefficients, and for very large switching times both $w_2$ and $w_3$ approach zero.   

The naive expectation that the pre-equilibrium dilution of the source eccentricity before $\tau_s$ should reduce the finally established anisotropic flow \cite{Kolb:2000sd,Heinz:2004pj} is borne out only if one completely ignores the position-momentum correlations created by the pre-equilibrium dynamics and starts the hydrodynamic stage with zero transverse flow. This is illustrated by the dashed lines with open symbols in Fig.~\ref{F11}.

Figure~\ref{F11} thus leads, in disagreement with the earlier statements made in Refs.~\cite{Kolb:2000sd, Heinz:2004pj}, to the (revised) conclusion that elliptic and higher-order anisotropic flow measurements alone cannot put a tight upper limit on the duration of the pre-equilibrium stage. Fig.~\ref{F11} suggests that, as long as pre-equilibrium contributions to the final flow pattern are consistently accounted for, the final anisotropic flows are insensitive to how strongly the medium is coupled during the first 2\,fm/$c$ or so. Whether this stage is described hydrodynamically (very strong coupling) or by free-streaming partons (very weak coupling), one sees the same final flow anisotropy. This finding supports the idea of ``universal transverse flow'' during the earliest stages of the fireball evolution proposed by Pratt and Vredevoogd \cite{Vredevoogd:2008id}. In contrast, radial flow (which affects the slope and mean $p_\perp$ of the transverse momentum spectra) exhibits a strong and monotonic $\tau_s$ dependence already for small switching times that can be used much more effectively to put an upper limit on the thermalization time in heavy-ion collisions.

\section{Parameter optimization}
\label{parameterSearch}

In addition to the switching (or starting) time $\tau_s$, our hydrodynamical model has several other input parameters whose choice influences the final physical observables. In the preceding sections we only varied $\tau_s$, leaving these other parameters unchanged, in order to gain generic insights into which of the different observables at our disposal provide the strongest constraints on $\tau_s$. However, it is immediately obvious that there should be some sort of tradeoff between effectively weakening the interactions in the pre-equilibrium stage, say by lengthening the free-streaming period, and weakening the interactions during the later hydrodynamic stage, say by shortening the free-streaming stage and increasing the shear viscosity during the subsequent hydrodynamic evolution. The effects of changing the transition time between free-streaming and hydrodynaming evolution and of changing the shear viscosity during the hydrodynamic evolution are therefore entangled, and we should optimize both parameters simultaneously. Furthermore, since the slopes of the final spectra are controlled by a combination of the temperature and radial flow on the freeze-out hypersurface, and a change in viscosity affects the transverse pressure gradients and thus the radial flow, we should allow $T_\mathrm{dec}$ to vary together with $\eta/s$.

\begin{figure}[t!]
\hspace*{-2mm}
    \includegraphics[width=1.03\linewidth]{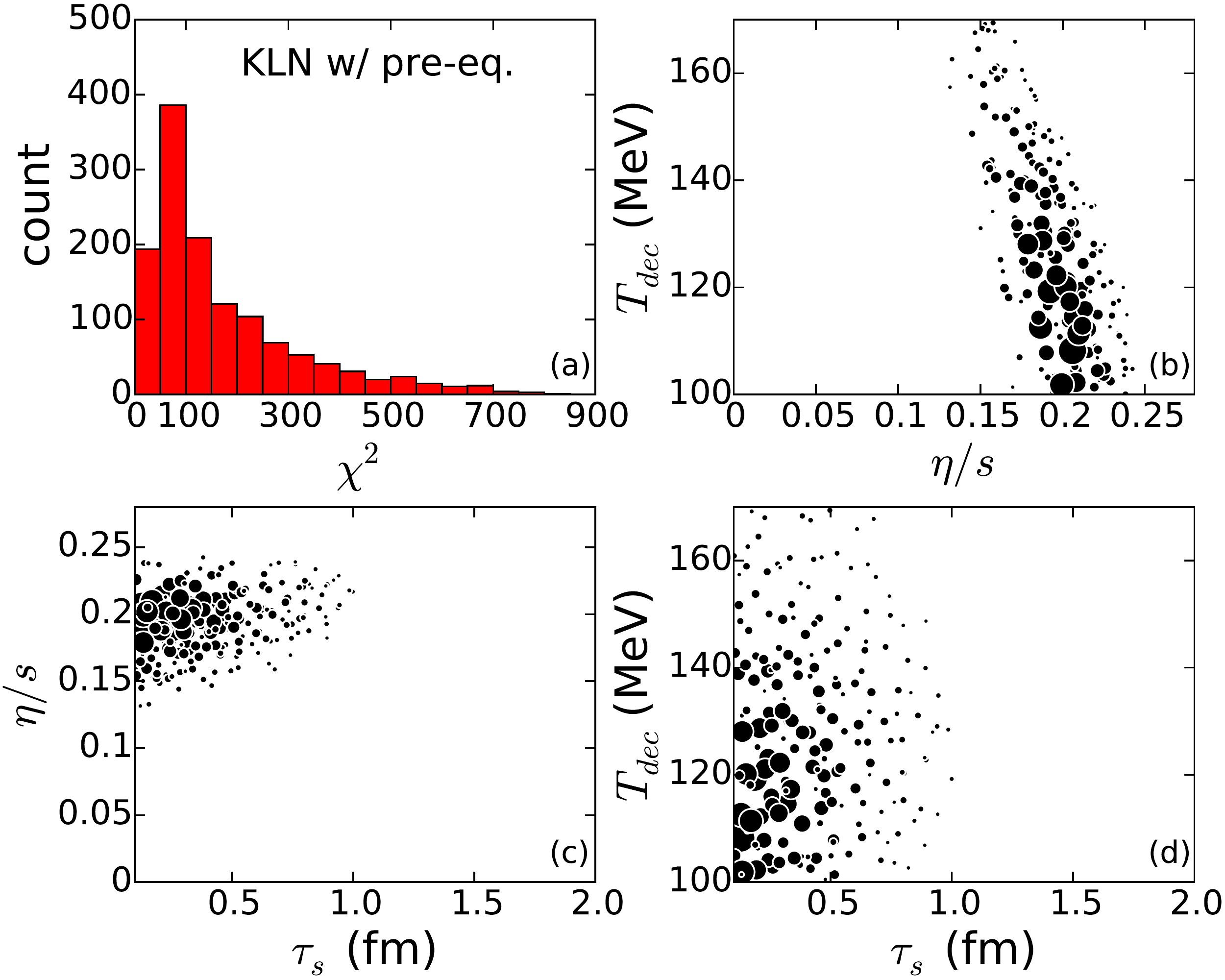}
    \caption{(Color online)
    Parameter search result for the KLN initial-state model, evolved in single-shot
    hydrodynamic mode following a free-streaming stage of 
    duration $\tau_s{-}\tau_0$. 
    (a) Histogram of the $\chi^2$ distribution. (b-d) 2-dimensional projections of 
    those parameter triplets $(\tau_s,\eta/s,T_\mathrm{dec}$) corresponding to 
    $\chi^2{\,<\,}50$. The size of the rings around their positions increases with 
    decreasing $\chi^2$, i.e. with increasing fit quality.}
    \label{F12}
\end{figure}
    
\begin{figure}[!t]
      \includegraphics[width=0.9\linewidth]{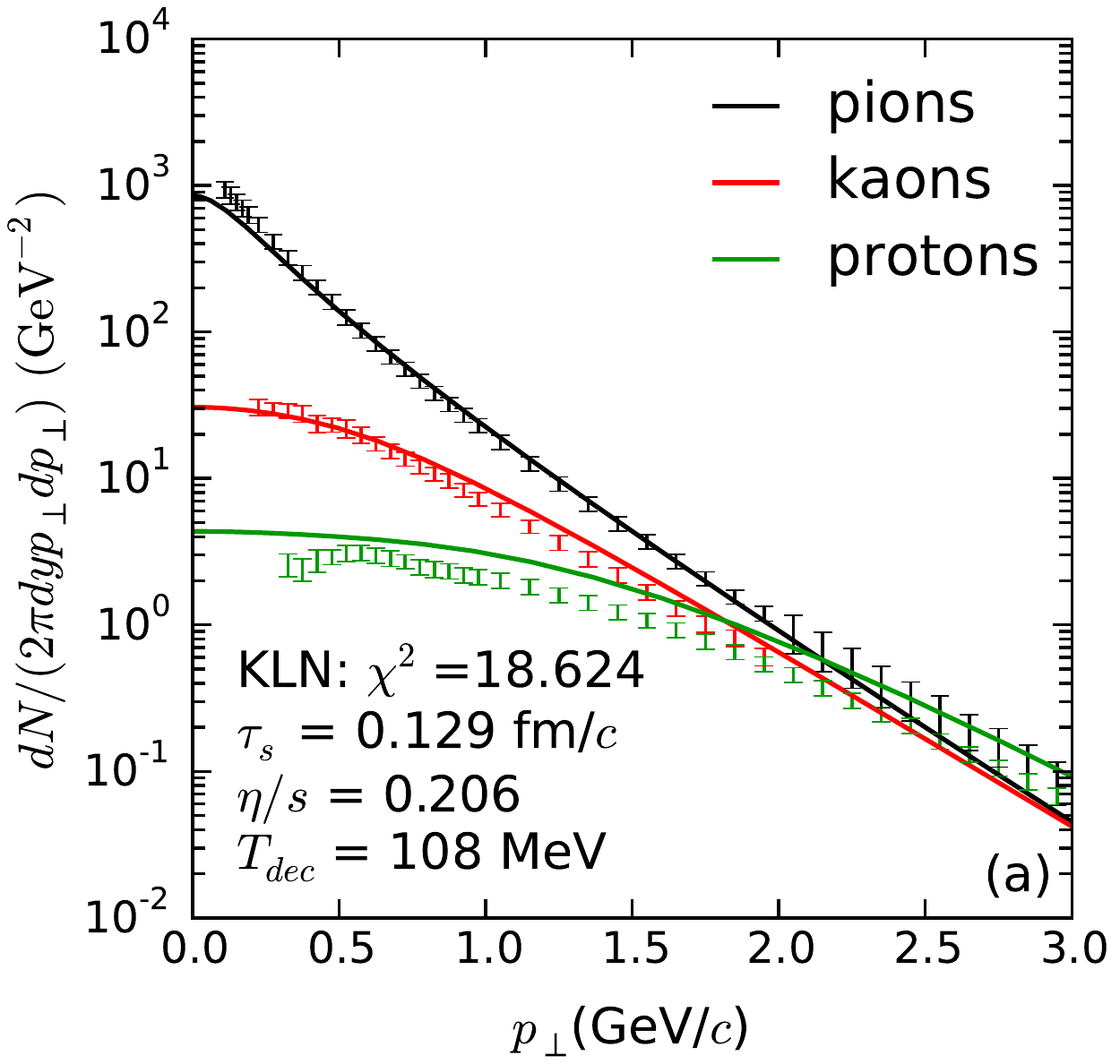}\\
      \hspace*{6mm}
      \includegraphics[width=0.82\linewidth]{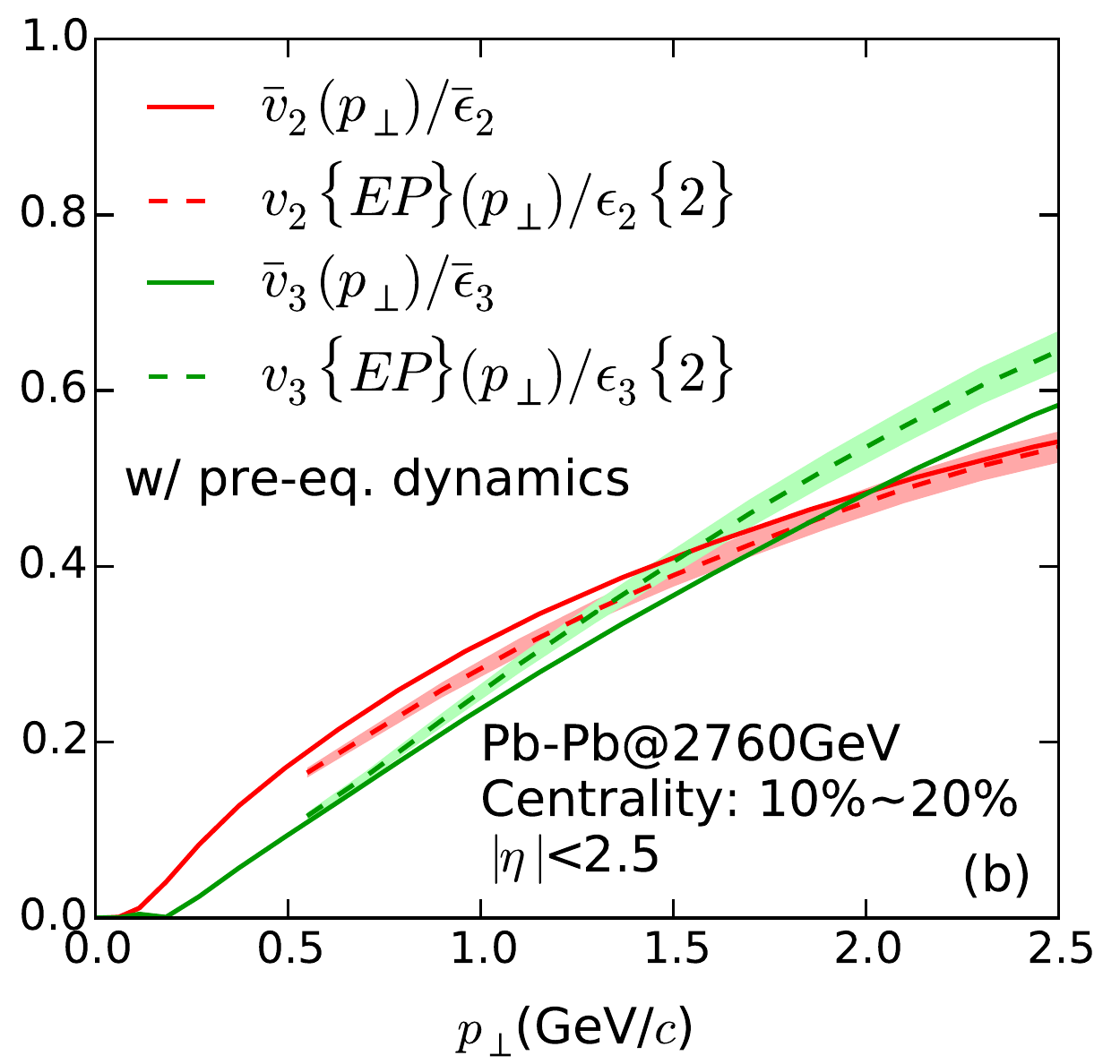}
      \caption{Comparison with experimental data of single-shot hydrodynamic model 
      predictions with the best fit parameter combination (smallest $\chi^2$ value) for 
      KLN-initialized simulations with pre-equilibrium free-streaming dynamics, for 
      $2.76\,A$\,TeV Pb+Pb collisions at 10\%--20\% centrality. The 
      best-fit parameter values are listed in panel (a). (a) Transverse momentum spectra 
      for $\pi^+$, $K^+$ and protons,  compared with ALICE data \cite{Preghenella:2012eu}.
      (b) Eccentricity-scaled $p_\perp$-differential elliptic and triangular flow anisotropies 
      for charged hadrons, compared with ATLAS data \cite{ATLAS:2012at}. Single-shot
      hydrodynamic simulations predict the ensemble averaged flow $\bar{v}_n$ which
      was scaled by the corresponding ensemble-averaged eccentricity $\bar{\epsilon}_n$ 
      (solid lines). The ATLAS anisotropic flow data are measured with the event-plane 
      method and are thus affected by the variance of the event-by-event $v_n$ distribution.
      They were fitted with a smooth curve and scaled by the rms eccentricity 
      $\epsilon_n\{2\}$ (dashed lines). The shaded area around the dashed lines represents
      the experimental error of the ATLAS $v_n\{\mathrm{EP}\}$ measurements 
      \cite{ATLAS:2012at}.
        \label{F13}
      }
\end{figure}

\begin{figure}[!t]
\hspace*{-2mm}
    \includegraphics[width=1.03\linewidth]{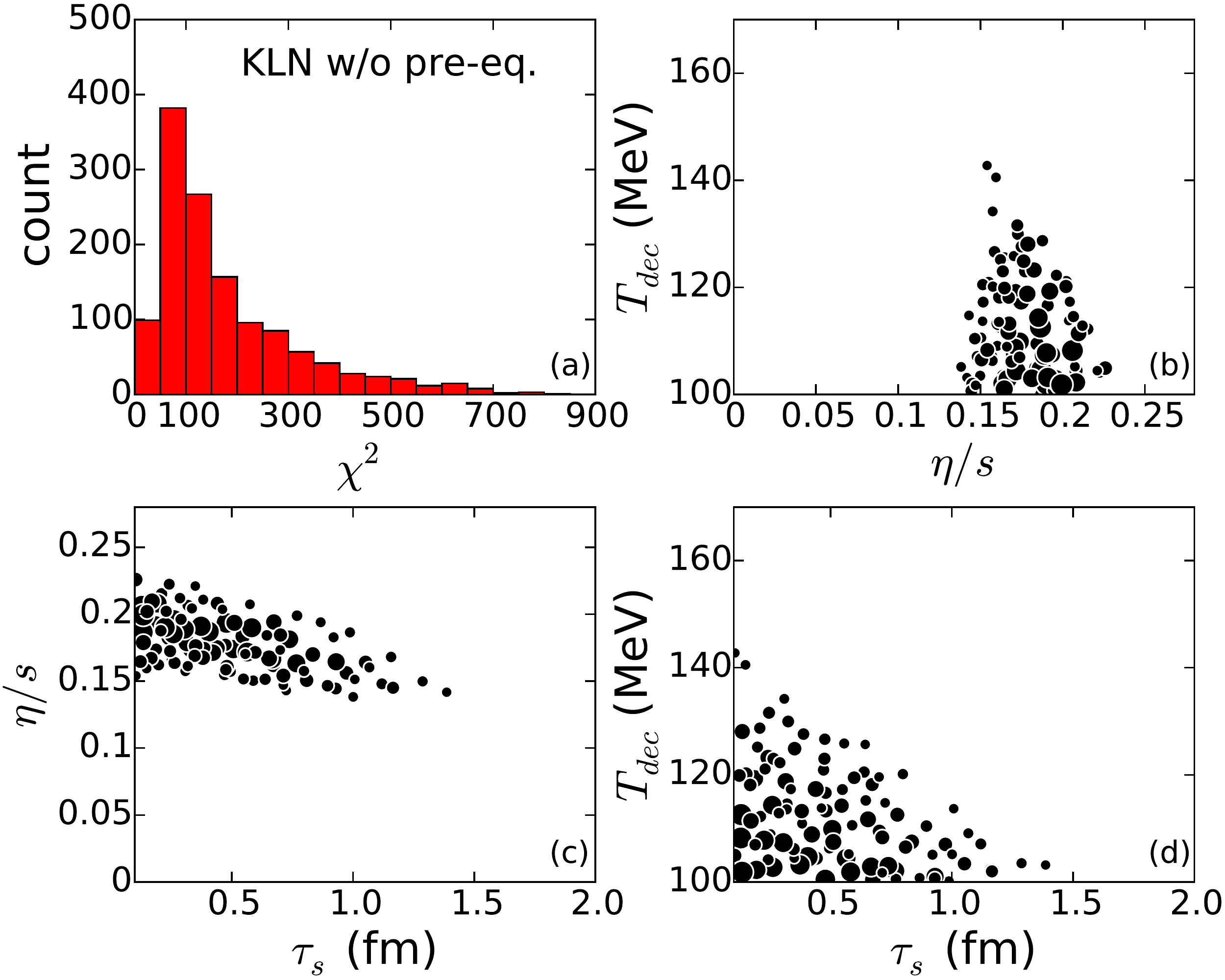}
    \caption{(Color online)
    Same as Fig.~\ref{F12}, but for single-shot hydrodynamic simulations 
    starting at $\tau_s$ without preceding pre-equilibrium stage.
    \label{F14}
    }
\end{figure} 
    
\begin{figure}[!t]              
     \includegraphics[width=0.9\linewidth]{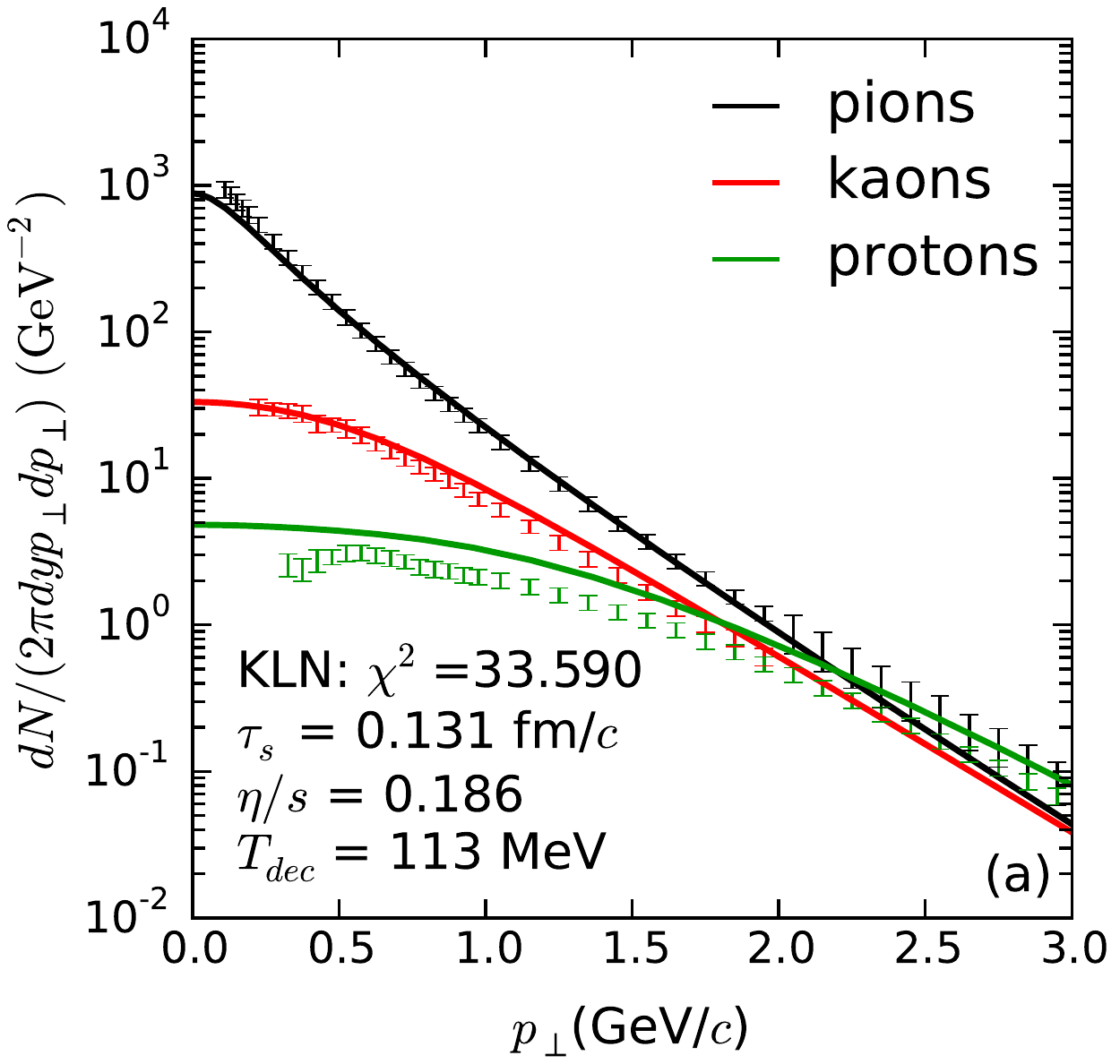}\\
     \hspace*{6mm}
     \includegraphics[width=0.82\linewidth]{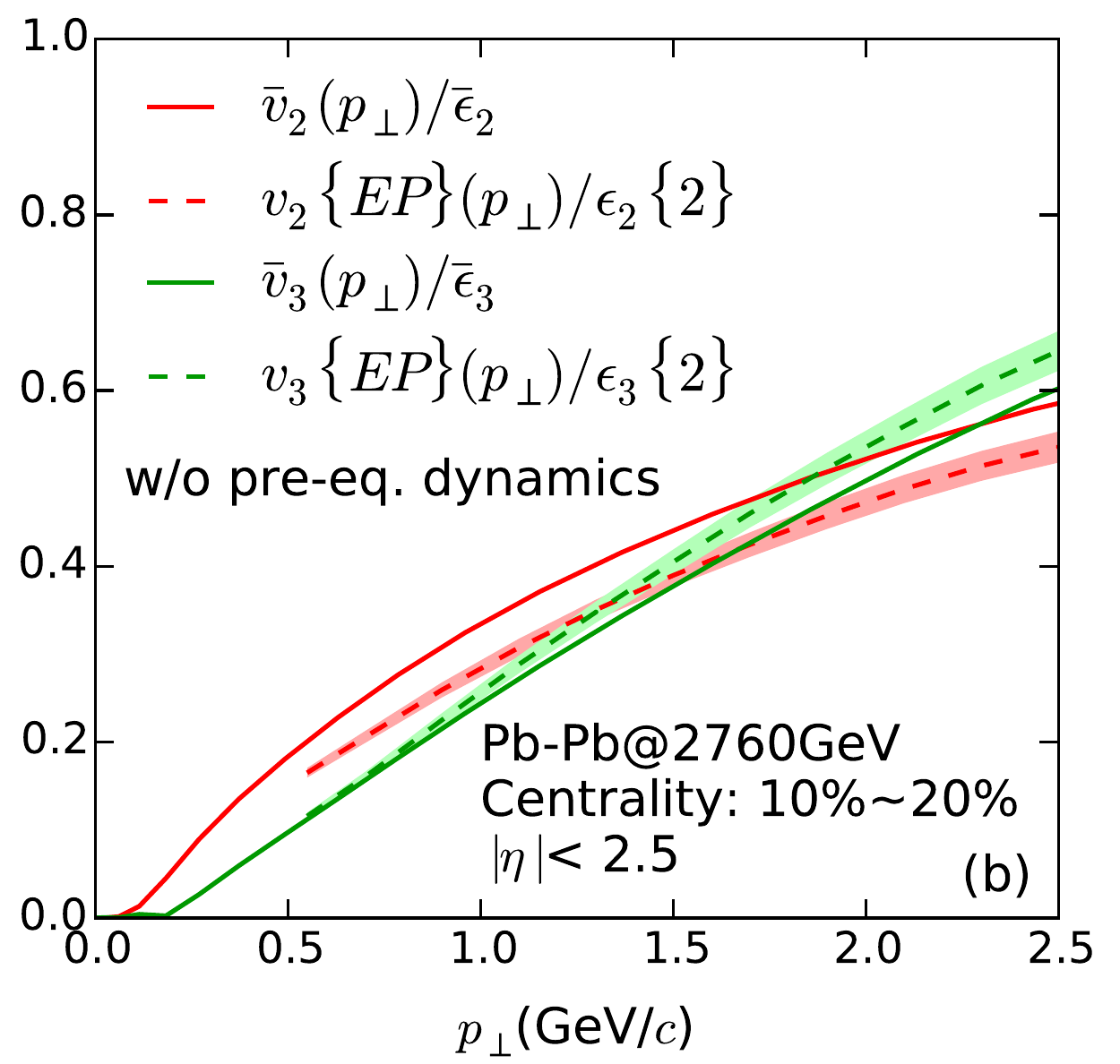}\\
      \caption{
      Same as Fig.~\ref{F13}, but for single-shot hydrodynamic simulations 
      starting at $\tau_s$ without preceding pre-equilibrium stage.
        \label{F15}      
      }     
\end{figure}

Here we report on a study where we allow $\tau_s$, $\eta/s$ and $T_\mathrm{dec}$ to vary simultaneously, trying to find the best combination by comparing the model predictions for $v_2^\mathrm{ch}$, $v_3^\mathrm{ch}$, $\langle p_\perp\rangle_{\pi^+}$, $\langle p_\perp\rangle_{K^+}$, and $\langle p_\perp\rangle_{p}$ for 2.76\,$A$\,TeV Pb+Pb collisions of 10--20\% centrality with experimental data from the ALICE \cite{Abelev:2013vea} and ATLAS \cite{Aad:2013xma} collaborations and minimizing the value of $\chi^2$. The values of the experimental measurements for these observables are summarized in Table~\ref{T1}.

\begin{table}[!h]
\centering
\begin{tabular}{lcc}
\hline \hline 
$\langle v_2^{\mathrm{ch}}\rangle$   &$0.0782 \pm 0.0019$ \\ 
$\langle v_3^{\mathrm{ch}}\rangle$   &$0.0316 \pm 0.0008$  \\ 
$\langle\scaperp{p}\rangle_{\pi^+}$ (GeV/$c$) & $0.517 \pm 0.017$  \\ 
$\langle\scaperp{p}\rangle_{K^+}$   (GeV/$c$) & $0.871 \pm 0.027$   \\ 
 $\langle\scaperp{p}\rangle_{p}$    (GeV/$c$) & $1.311\pm 0.034$ \\  
\hline\hline
\end{tabular}
\caption{Experimental data for the five hadronic observables from 2.76\,$A$\,TeV Pb+Pb 
              collisions of 10--20\% centrality that were considered in our fit. The mean 
              $p_\perp$-integrated elliptic and triangular flow values for charged hadrons were 
              measured by ATLAS \cite{Aad:2013xma}, the mean transverse momenta for positively
              charged pions, kaons and protons (extrapolated to the full $p_\perp$ range) by
              the ALICE Collaboration \cite{Abelev:2013vea}.               
              \label{T1}}
\end{table}

A somewhat more ambitious fit with five hydrodynamic model parameters and three experimental observables (charged multiplicity, $\langle v_2^\mathrm{ch}\rangle$ and $\langle v_3^\mathrm{ch}\rangle$) at six collision centralities each (i.e. alltogether 18 observables), including a hadronic afterburner but no pre-equilibrium dynamics, was recently reported in \cite{Bernhard:2015hxa}. We perform simulations both with and without pre-equilibrium dynamics, in order to assess its impact on the best-fit values for the other model parameters. Our simulations are done in single-shot mode with smooth ensemble-averaged initial density profiles, not in event-by-event mode with fluctuating initial profiles as the work reported in \cite{Bernhard:2015hxa}. Since we used data on the mean elliptic and triangular flows of charged hadrons, obtained by ATLAS \cite{Aad:2013xma} from their full reconstructed event-by-event probability distributions, instead of their rms values that were used in \cite{Bernhard:2015hxa} and which are affected by the variance of their event-by-event fluctuations, the added numerical cost of event-by-event hydrodynamic simulations could be avoided in our analysis.

The parameter space is explored by Latin hypercube sampling \cite{lhs,lhsPackage}, which is a statistical sampling method for optimizing the selection of parameter combinations from a high-dimensional parameter space. Because our parameter space is 3-dimensional, each point drawn from this space by the Latin hypercube method contains three components. In order to efficiently span the parameter space, no two points in the sample share the same value for any of the three parameters. We checked that, after projecting the sampled triplets onto any of the three components, the distribution of that component was uniform.

\begin{figure}[t!]
\hspace*{-2mm}
    \includegraphics[width=1.03\linewidth]{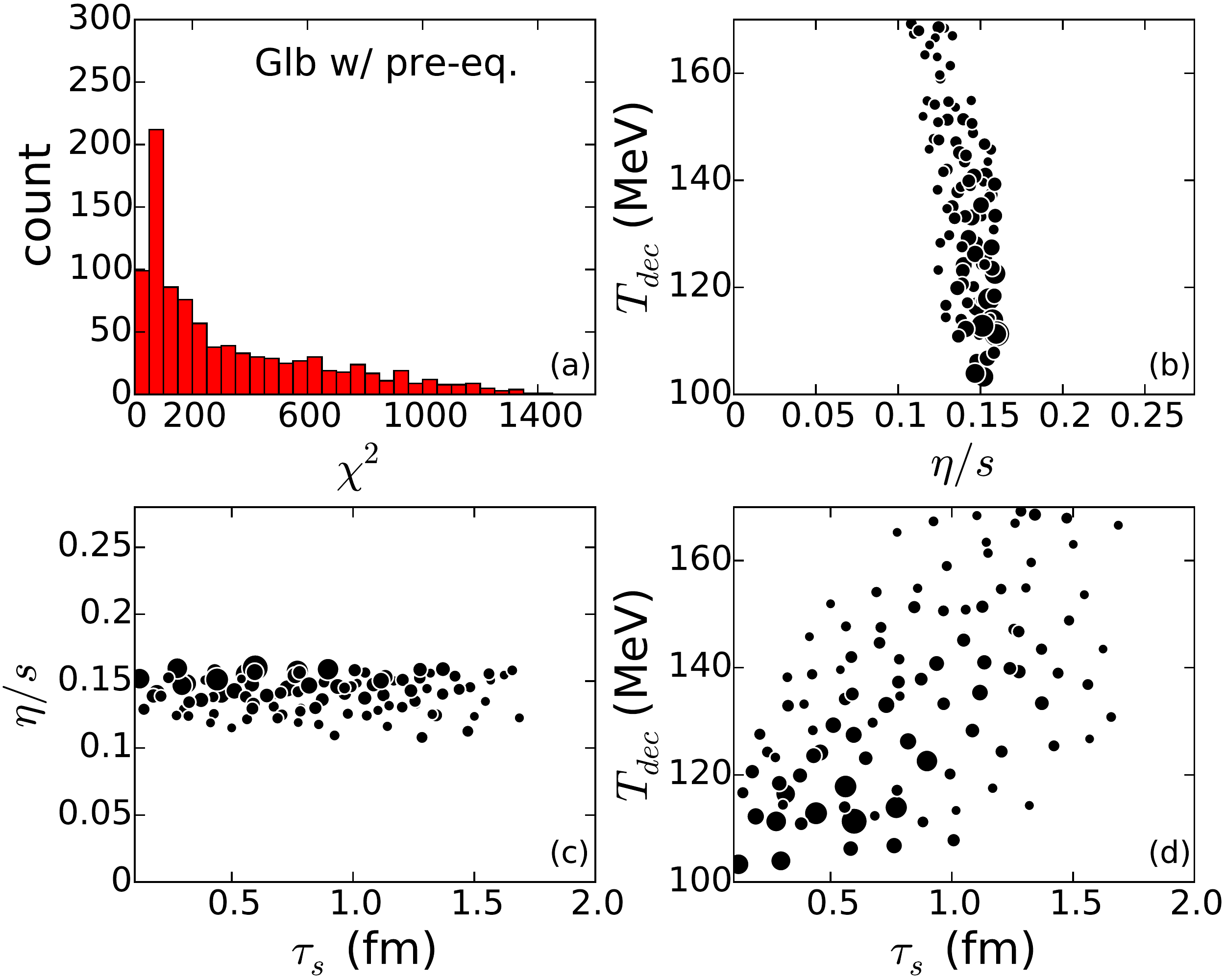}
    \caption{(Color online)
    Same as Fig.~\ref{F12}, but for MC-Glauber initial conditions, free-streamed
    to $\tau_s$ and then evolved hydrodynamically.
    \label{F16}
    }
\end{figure}

\begin{figure}[t!]
      \includegraphics[width=0.9\linewidth]{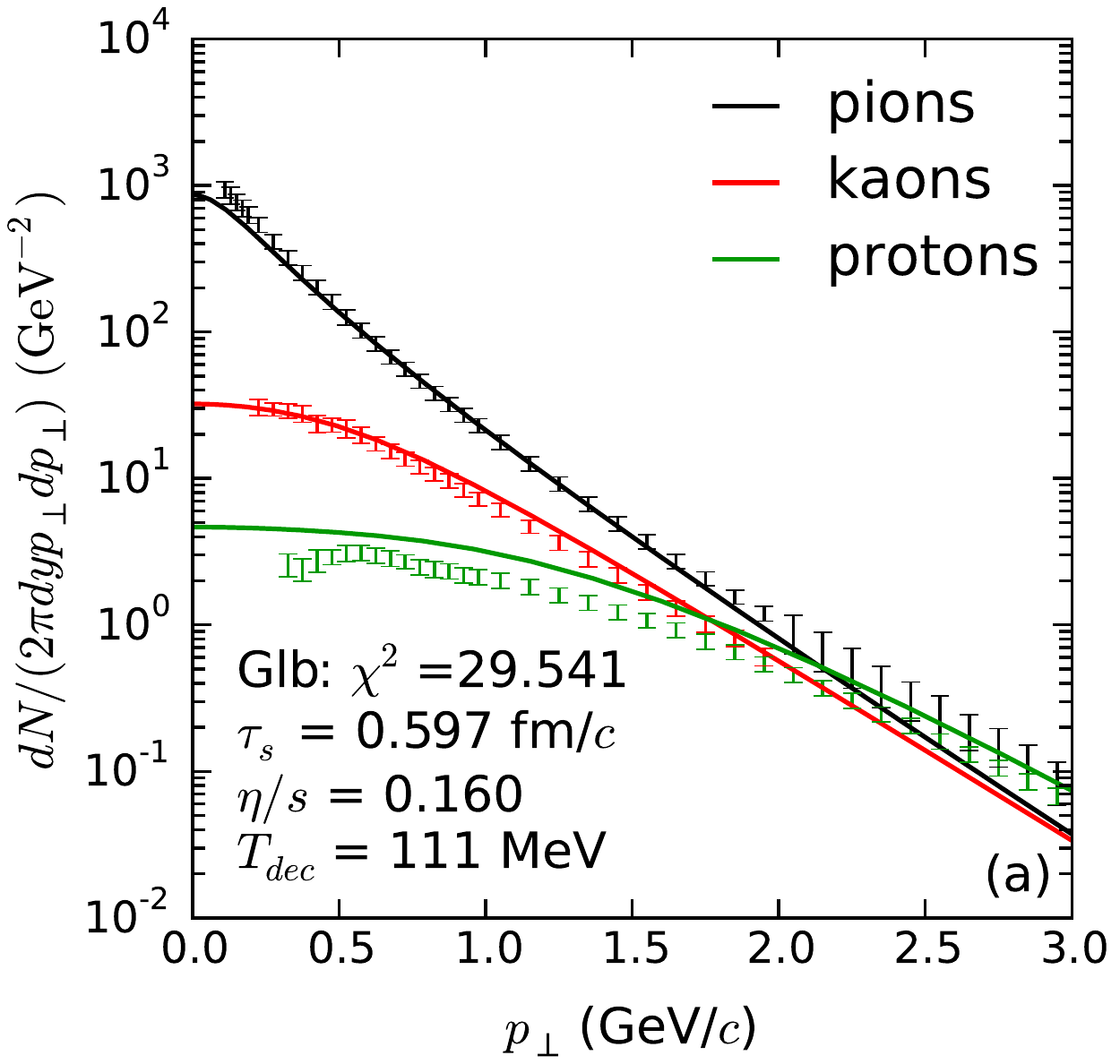}\\
      \hspace*{6mm}
      \includegraphics[width=0.82\linewidth]{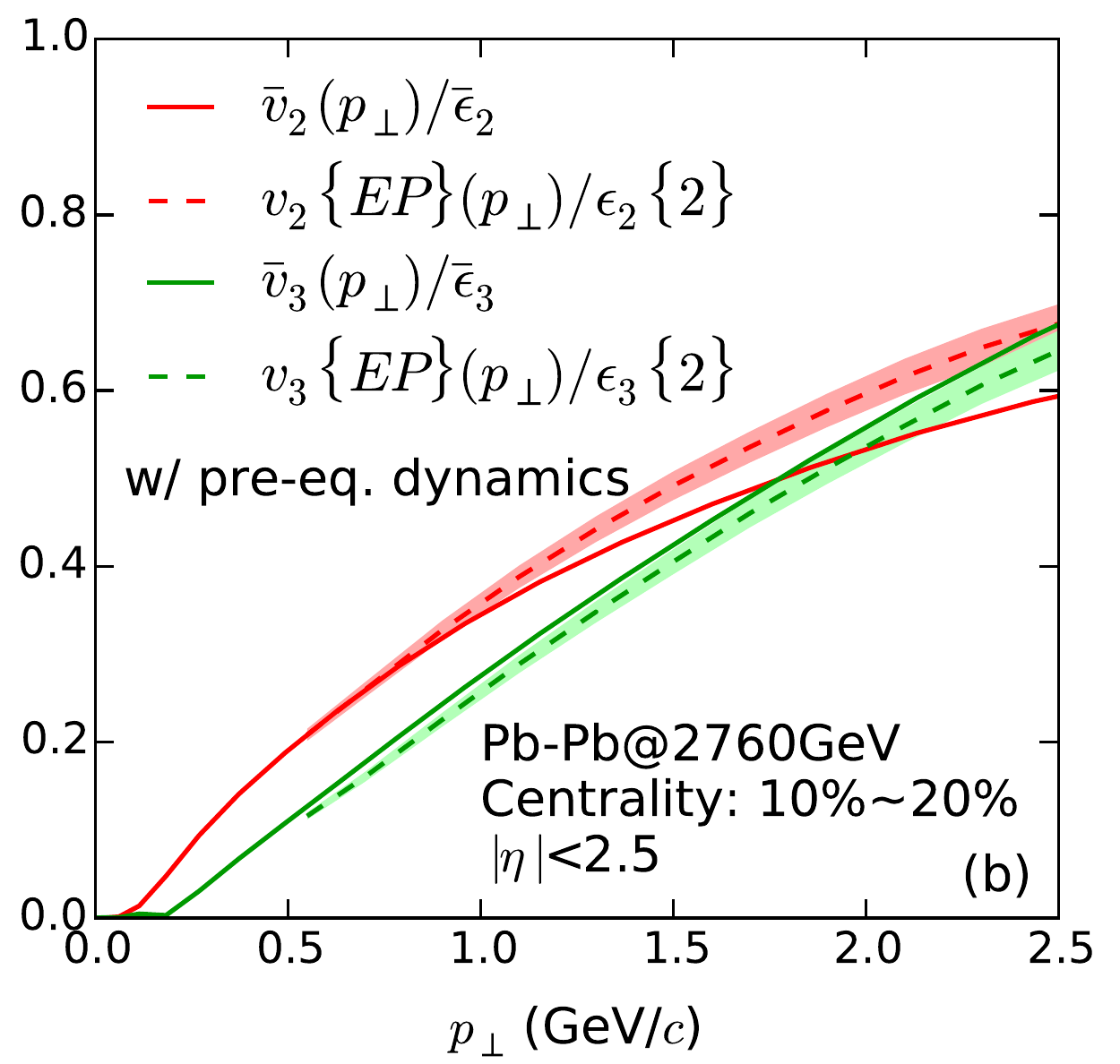}
      \caption{Same as Fig.~\ref{F13}, but for MC-Glauber initial conditions, free-streamed
        to $\tau_s$ and then evolved hydrodynamically.
        \label{F17}
       }
\end{figure}

After running the hydrodynamic simulation for a given triplet of parameters from the Latin hypercube sample, the quality of the description obtained with this parameter set is estimated by computing the $\chi^2$ of the resulting fit of the selected experimental data:
\begin{eqnarray}
\chi^2 = \sum_i  \frac{(O_i - E_i)^2}{\sigma_i^2}.
\label{chisqaure}
\end{eqnarray}
Here the sum runs over the five observables, $E_i$ is the value and $\sigma_i$ the combined statistical and systematic error of the experimental measurement of the observable, and $O_i(\tau_s,\eta/s,T_\mathrm{dec})$ is the value of the observable from the simulation with parameter set $(\tau_s,\eta/s,T_\mathrm{dec})$. Since we are using single-shot hydrodynamics with an ensemble-averaged initial profile, $O_i$ has no statistical error. Assuming that the five chosen observables are uncorrelated, with three fit parameters we have two statistical degrees of freedom, and a good fit of the data should thus have $\chi^2/2{\,\simeq\,}1$. The best fits we have been able to achieve with the three selected model parameters have $\chi^2/2 \sim 9-15$ (see Table~\ref{T3}). This suggests that at least one additional physical mechanism not captured by these three model parameters may play an important role in describing the chosen observables. This could be, for example, that hydrodynamics breaks down during the final hadronic stage of the fireball expansion and needs to be replaced there by a microscopic model for hadronic rescattering \cite{Bernhard:2015hxa,Song:2010aq}. We have not tested this hypothesis.

\begin{figure}[t!]
 \hspace*{-2mm}
    \includegraphics[width=1.03\linewidth]{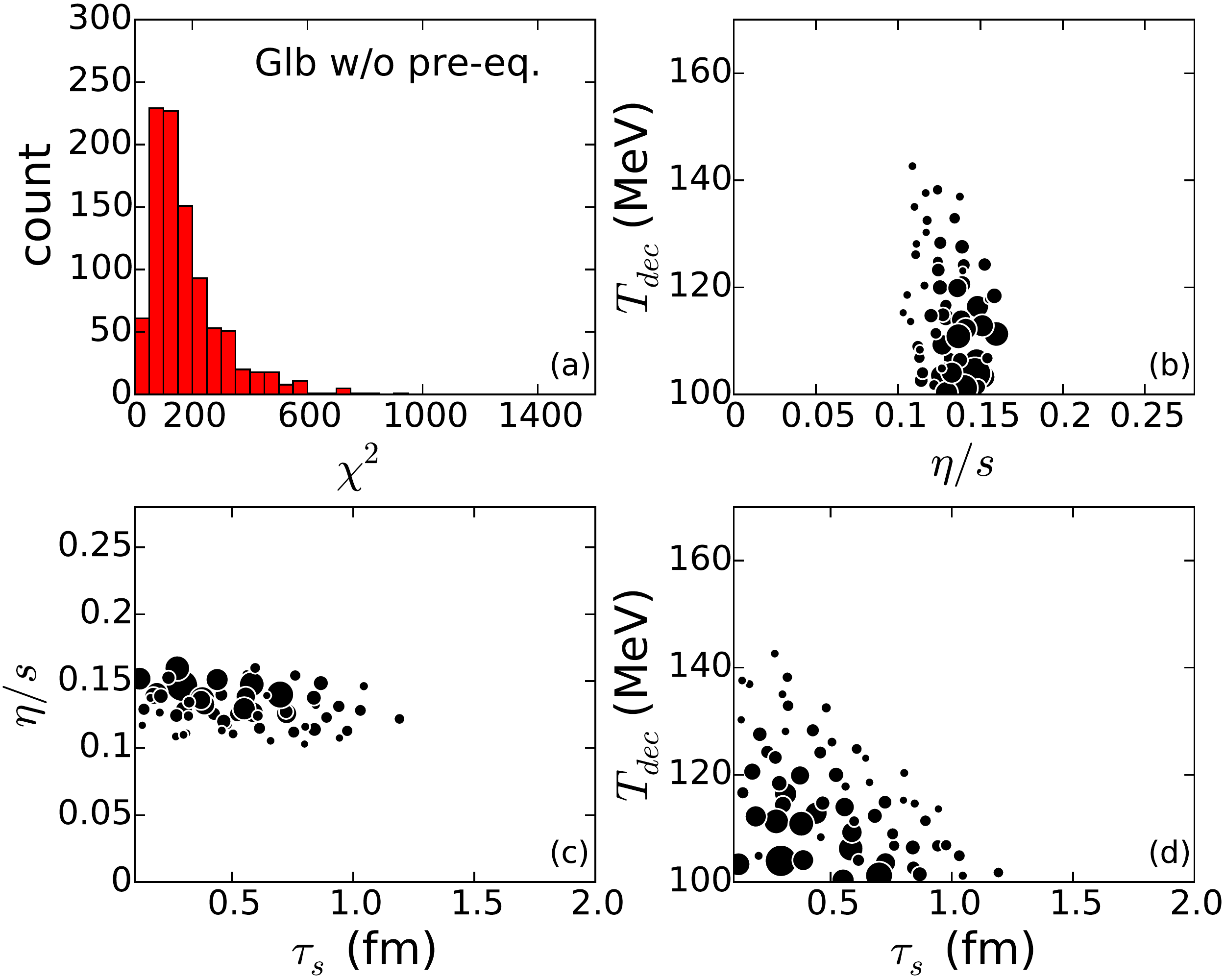}
    \caption{(Color online)
    Same as Fig.~\ref{F16}, but for single-shot hydrodynamic simulations 
    starting at $\tau_s$ without preceding pre-equilibrium stage. 
    \label{F18}
    }
\end{figure}
                  
\begin{figure}[t!]                  
     \includegraphics[width=0.9\linewidth]{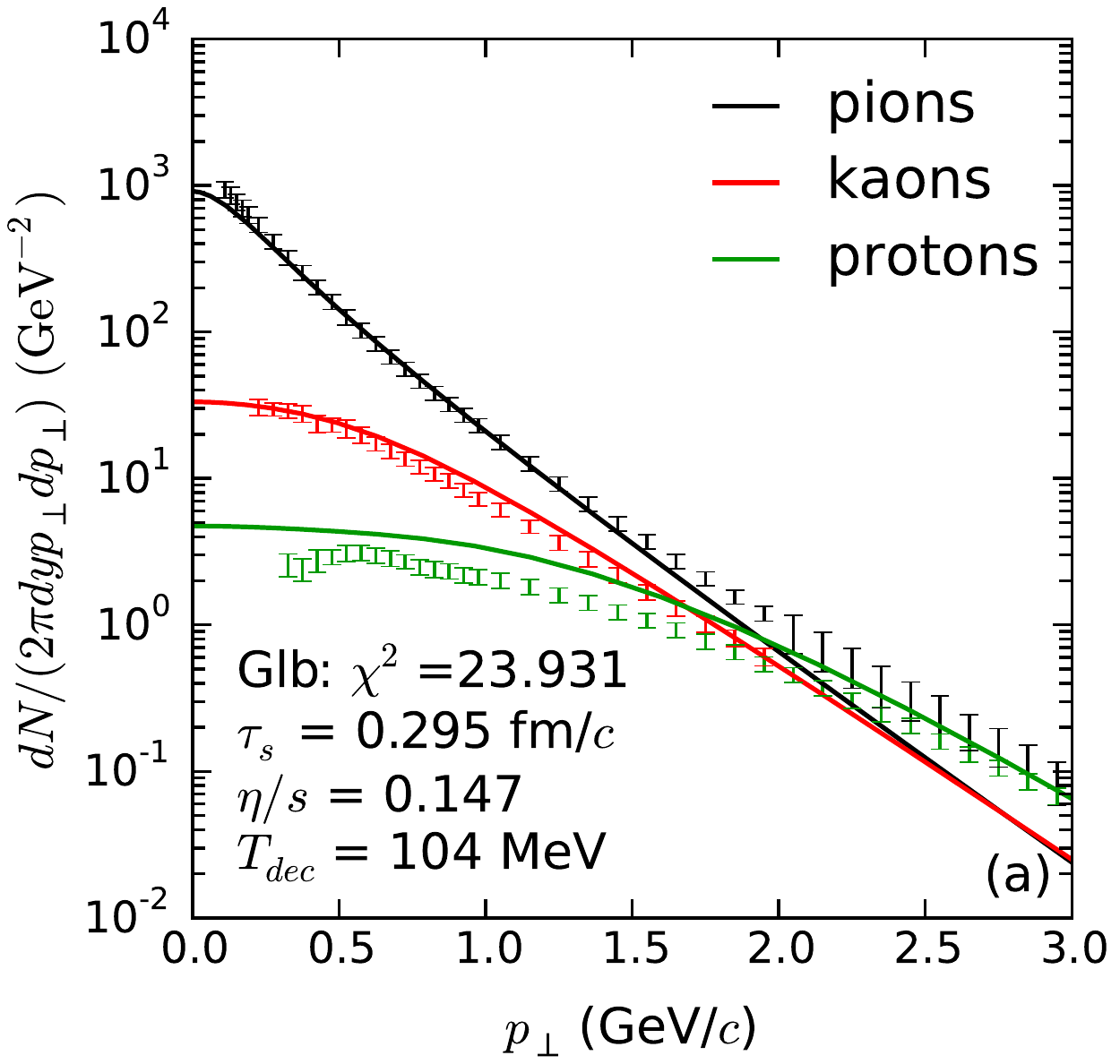}\\
     \hspace*{6mm}
     \includegraphics[width=0.82\linewidth]{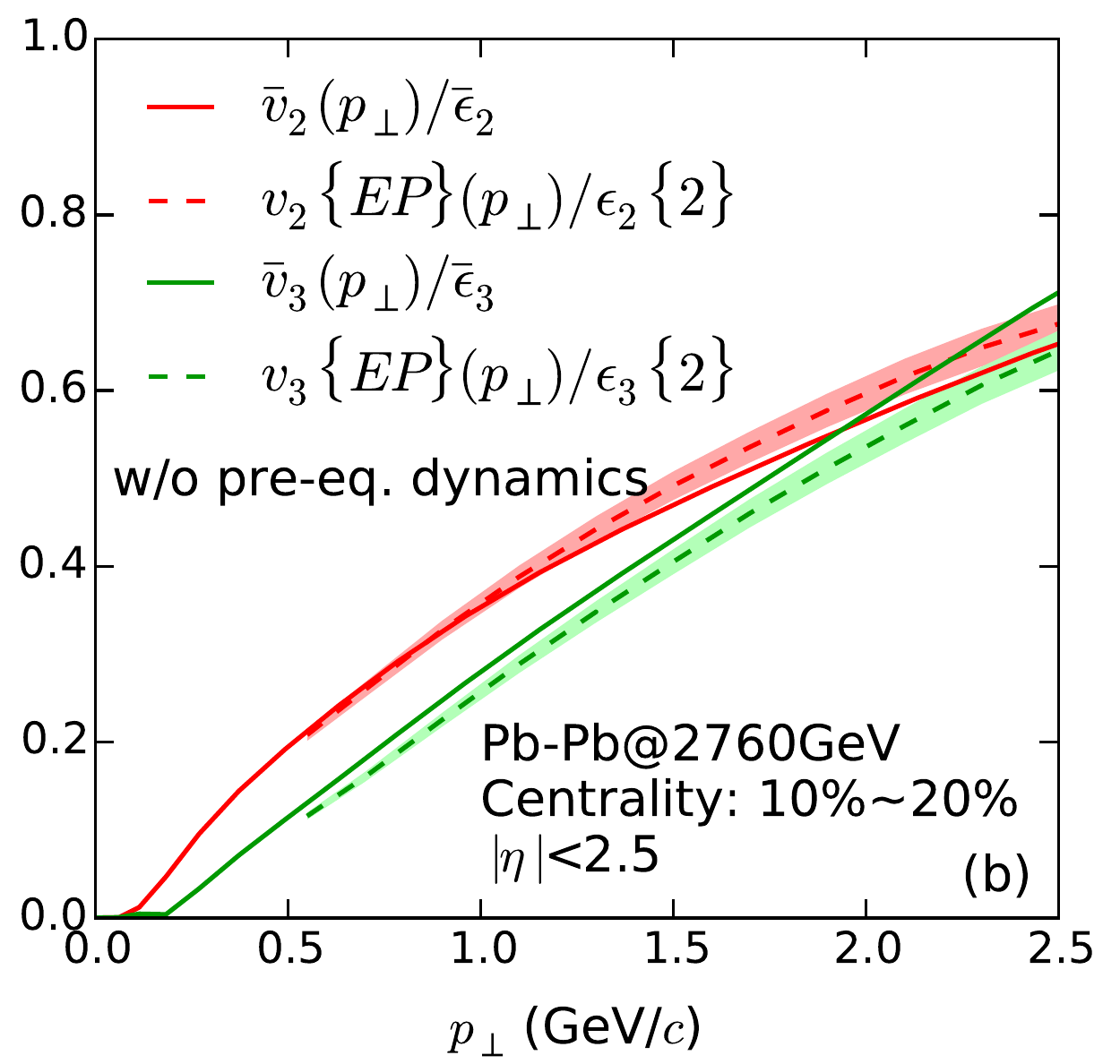}\\
      \caption{
      Same as Fig.~\ref{F17}, but for single-shot hydrodynamic simulations 
      starting at $\tau_s$ without preceding pre-equilibrium stage.  
        \label{F19}
     }
\end{figure}

For the KLN initial-state model we drew a Latin hypercube sample of 1300 points covering the following parameter ranges: $0.1\,\mathrm{fm}/c{\,<\,}\tau_s{\,<\,}1.4$\,fm/$c$, $0.08{\,<\,}\eta/s{\,<\,}0.28$, and 100\,MeV${\,<\,}T_\mathrm{dec}{\,<\,}170$\,MeV. For the Glauber initial-state model we sampled
950 points in the range $0.1\,\mathrm{fm}/c{\,<\,}\tau_s{\,<\,}2$\,fm/$c$, $0{\,<\,}\eta/s{\,<\,}0.16$, and 100\,MeV${\,<\,}T_\mathrm{dec}{\,<\,}170$\,MeV.%
\footnote{%
     In both cases the upper end of the explored range for $\tau_s$ is small 
     enough that the particle and energy losses through the corona studied
     in Sec.~\ref{fsIssues} can be ignored.}
Figs.~\ref{F12}, \ref{F14}, \ref{F16} and \ref{F18} show the corresponding $\chi^2$ distributions for simulations with KLN and Glauber initial conditions
with and without a free-streaming pre-equilibrium stage, respectively (see captions). The best-fit values for the three model parameters in each of the four cases, and the predictions the model makes for these best-fit parameter values for the observables listed in Table~\ref{T1}, are summarized in Tables~\ref{T2} and \ref{T3}. 
%
\begin{table}[!b]
\centering
\begin{tabular}{ccccc}
\hline \hline 
Model  &  Pre-eq.  &  $\tau_s$ (fm/$c$) &  $\eta/s$  &  $T_{dec}$ (MeV) \\ \hline \hline
MC-KLN  & Yes & 0.129 & 0.206 & 108\\
MC-KLN  &  No & 0.131 & 0.186 &  113\\
MC-Glb  & Yes  & 0.597 & 0.160 &  111\\
MC-Glb  & No   & 0.295 & 0.147 &  104\\
\hline\hline
\end{tabular}
\caption{Best-fit parameters from a parameter search for four different types of simulations. 
             \label{T2}}
\end{table}
%
\begin{table}[!b]
\centering
\begin{tabular}{lcccc}
\hline \hline 
Model                          &  MC-KLN  &  MC-KLN &  MC-Glb  &  MC-Glb \\ 
Pre-eq.                        & Yes & No & Yes & No\\ \hline\hline
$\bar v_2^{\mathrm{ch}}$  &  0.083  & 0.088 & 0.070 & 0.071 \\
$\bar v_3^{\mathrm{ch}}$  & 0.030   & 0.030 & 0.034 & 0.034 \\
$\langle\scaperp{p}\rangle_{\pi^+}$ (GeV/$c$)  & 0.550 & 0.545 & 0.539 & 0.518 \\
$\langle\scaperp{p}\rangle_{K^+}$   (GeV/$c$)  & 0.900 & 0.877 & 0.869 & 0.852 \\
$\langle\scaperp{p}\rangle_{p}$     (GeV/$c$)  & 1.349 & 1.302 & 1.293 & 1.279 \\
$\chi^2$                           & 18.624 & 33.590 & 29.541 & 23.931 \\
\hline\hline
\end{tabular}
\caption{Model predictions for the five observables listed in Table~\ref{T1} with the 
              best-fit parameters listed in Table~\ref{T2}.
              \label{T3}}
\end{table}
%
Panels (b-c) in Figs.~\ref{F12}, \ref{F14}, \ref{F16} and \ref{F18} show scatter plots of all simulations with a total $\chi^2{\,<\,}50$ in each of the three 2-dimensional projections of the 3-dimensional parameter space. The size of each blob indicates the quality of the description of the data achieved with the corresponding parameter set: The larger the blob, the smaller the total $\chi^2$ and the better the fit.

Figures~\ref{F13}, \ref{F15}, \ref{F17} and \ref{F19} show the $p_\perp$ distributions of pions, kaons and protons (a) and of the charged hadron elliptic and triangular flows predicted by the simulations with the best-fit parameter sets, for each of the four simulation modes listed in Tables~\ref{T2}, \ref{T3}. We use the equation of state s95p-PCE \cite{Huovinen:2009yb,Shen:2010uy} which assumes chemical freeze-out of hadron abundances at $T_\mathrm{chem}\eq165$\,MeV and therefore overpredicts the measured proton yields by about 50\% (without the hadronic afterburner, our hydrodynamic approach cannot account for baryon-anti-baryon annihilation during the final hadronic rescattering stage, which is required to reproduce the experimental yields \cite{Song:2013qma}). For the protons one should therefore ignore the normalization of the $p_\perp$ spectra and focus instead on their shape. Clearly, the description of the experimental data in Figs.~\ref{F13}, \ref{F15}, \ref{F17} and \ref{F19} (of which only the lowest non-trivial moments were used in the fit) is not perfect, but of similar quality as most other, less systematic parameter fits published in the literature. 

At first sight, though, there appears to be one exception: As seen in Fig.~\ref{F19}b, the MC-Glauber model {\em without} pre-equilibrium dynamics appears to provide a qualitatively better simultaneous description of the differential charged hadron elliptic and triangular flows than the other approaches. (In particular, the KLN model has troubles to describe these two observables simultaneously, as has been noted before \cite{Qiu:2011hf}.) However, the overall $\chi^2$ of this ``Glauber without pre-equilibrium'' fit is worse than that for the ``KLN with pre-equilibrium'' fit, due to a larger discrepancy with the data of the predicted mean $p_\perp$ for kaons and pions. Fig.~\ref{F19}a also shows that the best ``Glauber without pre-equilibrium'' simulation does not describe the slope of the pion spectrum quite as well as the best ``KLN with pre-equilibrium'' fit in Fig.~\ref{F13}a. This clearly detracts from the apparently much better description of the $p_\perp$-differential $v_{2,3}^\mathrm{ch}(p_\perp)$ in Fig.~\ref{F19}b compared to Fig.~\ref{F13}b which may simply reflect an incorrect weighting of regions of high and low $p_\perp$ in the best-fit ``Glauber without pre-equilibrium'' simulation.

Let us discuss a few other trends that are visible in Figs~\ref{F12}-\ref{F19}. First, panels b and c in Figs.~\ref{F12}, \ref{F14}, \ref{F16} and \ref{F18}, as well as Table~\ref{T2}, show that the inclusion of pre-equilibrium dynamics puts some upward pressure on the best-fit value of $\eta/s$, for both types of initial conditions. The effect is not large (of order 10\%) but appears to be significant. (We will not be absolutely certain of the significance of this observation until we have completed a full Markov Chain Monte Carlo (MCMC) simulation of the posterior model parameter distributions \cite{Bernhard:2015hxa, Pratt:2015zsa} which is beyond the scope of this paper.) Second, by looking at the spread of the blobs shown in panel b of Figs.~\ref{F12}, \ref{F14}, \ref{F16} and \ref{F18}, we see a reduced sensitivity of the fit quality to the decoupling temperature when pre-equilibrium dynamics is accounted for in the simulations: the same upper limit of 50 for the total $\chi^2$ allows for larger deviations of $T_\mathrm{dec}$ from the best-fit value if the simulations include a pre-equilibrium stage. Third, it can be seen from panel c in the same figures that allowing for pre-equilibrium build-up of flow reduces the probability of a good fit for switching times that significantly exceed the best-fit value in the KLN model but not in the Glauber model. This is consistent with Fig.~\ref{F10} where we observed that allowing for any appreciable delay $\tau_s$ of the hydrodynamic stage due to pre-equilibrium dynamics tends to cause the model to overpredict the mean $\scaperp{p}$ of the hadrons (especially the protons) for KLN initial profiles (panel a) but not for the Glauber model profiles (panel b). Indeed, for the Glauber model the neglect of pre-equilibrium flow puts downward pressure in $\tau_s$, because of the effect of missing radial flow on the mean hadron $\scaperp{p}$. One sees this in both Fig.~\ref{F10}b and when comparing panels c in Figs.~\ref{F16} and \ref{F18}. 

\section{Summary and conclusions}
\label{paperConclusion}

In this paper, we thoroughly investigated the effect of pre-equilibrium flow on heavy-ion collision observables. In order to estimate the maximum difference between simulations where the hydrodynamic stage was initiated with non-zero flow resulting from a preceding stage of pre-equilibrium evolution and others where all dynamical effects before the onset of hydrodynamic expansion were simply ignored (as has been common practice in many earlier studies), we here assumed the pre-equilibrium stage to be non-interacting, i.e. free-streaming. By adjusting the switching time $\tau_s$ between the free-streaming and hydrodynamic stages we can thus smoothly switch between a picture where the earliest stage of the collision is coupled infinitely weakly and one were it is coupled infinitely strongly.

The pre-equilibrium and hydrodynamic stages are Landau-matched, conserving energy and momentum and fully accounting for all components of the pre-equilibrium energy-momentum tensor with non-zero (and possibly large) dissipative pressure components (bulk and shear viscous pressures). The study presented here was done for zero bulk viscosity, i.e. the non-zero bulk viscous pressure generated by the Landau matching procedure was allowed to evolve dynamically to zero with a short microscopic relaxation time. The viscous shear stress at the beginning of the hydrodynmic stage was found to be large, corresponding to an inverse Reynolds number slightly above unity for early switching times and further increasing if one switches from free-streaming to hydrodynamics later. In the hydrodynamic stage, these large starting values for the shear stress decrease quickly on a short microscopic time scale, due to the assumed low shear viscosity $\eta/s$ of the hydrodynamic fluid.

On the way we had to solve the problem that extended free-streaming leads to a loss of a significant fraction of the energy of the expanding fireball because the volume of the ``corona'' of dilute matter that never thermalizes and therefore never becomes part of the hydrodynamic fluid increases. While we are not able to convert corona partons into hadrons we can account for their energy, so we can include them in anisotropic flow coefficients $w_n$ that characterize the azimuthal anisotropy of the energy flow even if we cannot account for their contributions to anisotropic particle flow $v_n$. We demonstrated that for small switching times were corona losses can be neglected, $w_n$ and $v_n$ closely track each other.

Our most important finding is that, contrary to traditional believes (previously also held by one of the present authors), the anisotropic flow coefficients are not very sensitive to the thermalization (or switching) time as long as the latter does not significantly exceeed about 2\,fm/$c$. The reason is that anisotropic flow not generated in the hydrodynamic stage, due a reduced initial fireball eccentricity and shortened hydrodynamic stage when $\tau_s$ is allowed to grow large, is almost precisely compensated for by anisotropies in the space-momentum correlations that develop during the free-streaming phase and manifest themselves as non-zero starting values for the momentum anisotropy of the energy-momentum tensor in the hydrodynamic stage. On the other hand, we found that an extended weakly coupled pre-equilibrium stage leads to stronger radial flow at the end of the hydrodynamic evolution, caused by large initial flow at the beginning of the hydrodynamic stage which is stronger than it would have been if the pre-equilibrium stage had been strongly coupled (i.e. describable by hydrodynamics). This is a result of the faster signal propagation speed in the pre-equilibrium stage which, in the free-streaming case, is given by the speed of light and thus exceeds the hydrodynamic speed of sound by almost a factor 2. This kick-start of the radial flow by pre-equilibrium evolution overcompensates the loss of radial flow generated during the hydrodynamic stage which is shortened when the pre-equilibrium stage lasts longer.

It thus turns out that the mean transverse momentum $\langle \scaperp{p}\rangle$ of the finally emitted hadrons, which (especially for the massive baryons) is strongly affected by the final radial flow of the fireball, provides a stronger upper limit on the thermalization time $\tau_s$ than the anisotropic flow coefficients.
This agrees with recent findings by Romatschke and collaborators \cite{Romatschke:2013re}. 

Of course, the switching time $\tau_s$ is only one of several model parameters affecting the final observables. In the last section we therefore performed an extended parameter search where we varied the switching time, decoupling temperature and specific shear viscosity of the fluid simultaneously, for two different initial state models and both runs with and without pre-equilibrium dynamics. While this does not exhaust the list of parameters and possibilities this exercise provides a more holistic picture of the effects of pre-equiibrium dynamics on final observables and on the values of medium parameters extracted from a comparison of the model predictions with experimental data. We found for both initial state models that accounting for pre-equilibrium dynamics slightly increases the optimal values for the specific shear viscosity $\eta/s$ extracted from mean $\scaperp{p}$ and anisotropic flow measurements and reduces the sensitivity of the observables to the decoupling temperature. However, the extracted limits for the thermalization time $\tau_s$ turned out to depend sensitively on the model for the initial density profiles.

\begin{acknowledgments}
U.~H. gratefully acknowledges early discussions with Tetsufumi Hirano in 2006 about a consistent Landau matching procedure to convert the pre-equilibrium stage to hydrodynamics and on the influence of a possible free-streaming early stage on the subsequent hydrodynamic evolution. These discussions first revealed several of the features described here, without leading, however, to a full understanding of their origins. J.~L. acknowledges support from the China Scholarship Council and fruitful discussions with Zhi~Qiu. The authors thank G.~S.~Denicol for kindly providing tables for the bulk viscous coefficients in Eq.~(\ref{f_bulk}). This work was supported by the U.S. Department of Energy, Office of Science, Office of Nuclear Physics under Awards No. \rm{DE-SC0004286} and (within the framework of the JET Collaboration) \rm{DE-SC0004104}.
\end{acknowledgments}



\begin{thebibliography}{99}

\bibitem{Kolb:2000sd} 
  P.~F.~Kolb, J.~Sollfrank and U.~Heinz,
  Phys.\ Rev.\ C {\bf 62}, 054909 (2000);
  P.~F.~Kolb and U.~Heinz,
  in {\it Quark-Gluon Plasma 3}, R.~C.~Hwa {\it et al.} (eds.) (World Scientific, Singapore, 2004), 
  pp. 634-714.

\bibitem{Broniowski:2008qk} 
  W.~Broniowski, W.~Florkowski, M.~Chojnacki and A.~Kisiel,
  Phys.\ Rev.\ C {\bf 80}, 034902 (2009).

\bibitem{Wu:2011yd} 
  B.~Wu and P.~Romatschke,
  Int.\ J.\ Mod.\ Phys.\ C {\bf 22}, 1317 (2011).

\bibitem{Heller:2012km} 
  M.~P.~Heller, D.~Mateos, W.~van der Schee and D.~Trancanelli,
  Phys.\ Rev.\ Lett.\  {\bf 108}, 191601 (2012).

\bibitem{Gelis:2013rba} 
  T.~Epelbaum and F.~Gelis,
  Phys.\ Rev.\ Lett.\  {\bf 111}, 232301 (2013).

\bibitem{Romatschke:2013re} 
  P.~Romatschke and J.~D.~Hogg,
  JHEP {\bf 1304}, 048 (2013);
  W.~van der Schee, P.~Romatschke and S.~Pratt,
  Phys.\ Rev.\ Lett.\  {\bf 111}, 222302 (2013).

\bibitem{Ryblewski:2013eja} 
  R.~Ryblewski and W.~Florkowski,
  Phys.\ Rev.\ D {\bf 88}, 034028 (2013).

\bibitem{Song:2007ux} 
  H.~Song and U.~Heinz,
  Phys.\ Rev.\ C {\bf 77}, 064901 (2008).

\bibitem{Shen:2014vra} 
  C.~Shen, Z.~Qiu, H.~Song, J.~Bernhard, S.~Bass and U.~Heinz,
  arXiv:1409.8164 [nucl-th].

\bibitem{Kharzeev:2001yq} 
  D.~Kharzeev, E.~Levin and M.~Nardi,
  Phys.\ Rev.\ C {\bf 71}, 054903 (2005).

\bibitem{Kharzeev:2004if} 
  D.~Kharzeev, E.~Levin and M.~Nardi,
  Nucl.\ Phys.\ A {\bf 747}, 609 (2005).

\bibitem{Alver:2008aq} 
  B.~Alver, M.~Baker, C.~Loizides and P.~Steinberg,
  arXiv:0805.4411 [nucl-ex].

\bibitem{Bjorken:1982qr} 
  J.~D.~Bjorken,
  Phys.\ Rev.\ D {\bf 27}, 140 (1983).
  
\bibitem{Baym:1984np} 
  G.~Baym,
  Phys.\ Lett.\ B {\bf 138}, 18 (1984).

\bibitem{Arnold:2014jva} 
  P.~Arnold, P.~Romatschke and W.~van der Schee,
  JHEP {\bf 1410}, 110 (2014).

\bibitem{Muronga:2001zk} 
  A.~Muronga,
  Phys.\ Rev.\ Lett.\  {\bf 88}, 062302 (2002)
  [Erratum-ibid.\  {\bf 89}, 159901 (2002)];
  and Phys.\ Rev.\ C {\bf 69}, 034903 (2004).

 \bibitem{Song:2009rh} 
  H.~Song and U.~Heinz,
  Phys.\ Rev.\ C {\bf 81}, 024905 (2010).

\bibitem{Drescher:2006ca} 
  H.-J.~Drescher and Y.~Nara,
  Phys.\ Rev.\ C {\bf 75}, 034905 (2007).

\bibitem{Alver:2010gr} 
  B.~Alver and G.~Roland,
  Phys.\ Rev.\ C {\bf 81}, 054905 (2010)
  [Erratum-ibid.\ C {\bf 82}, 039903 (2010)].

\bibitem{Alver:2010dn} 
  B.~H.~Alver, C.~Gombeaud, M.~Luzum and J.~Y.~Ollitrault,
  Phys.\ Rev.\ C {\bf 82}, 034913 (2010).

\bibitem{Betz:2008me} 
  B.~Betz, D.~Henkel and D.~H.~Rischke,
  Prog.\ Part.\ Nucl.\ Phys.\  {\bf 62}, 556 (2009).

\bibitem{Bazow:2013ifa} 
  D.~Bazow, U.~Heinz and M.~Strickland,
  Phys.\ Rev.\ C {\bf 90}, no. 5, 054910 (2014).

\bibitem{Heinz:2004pj} 
  U.~Heinz,
  AIP Conf.\ Proc.\  {\bf 739}, 163 (2005).

\bibitem{Heinz:2005zg} 
  U.~Heinz,
  in {\it Extreme QCD}, G. Aarts and S. Hands (eds.) (Swansea University, 2005), p. 3
  [arXiv:nucl-th/0512051].

\bibitem{Teaney:2009qa} 
  D.~A.~Teaney,
  in {\it Quark-Gluon Plasma 4}, R.~C.~Hwa and X.-N. Wang (eds.) 
  (World Scientific, Singapore, 2010), p. 207 [arXiv:0905.2433 [nucl-th]].

\bibitem{Cooper:1974mv} 
  F.~Cooper and G.~Frye,
  Phys.\ Rev.\ D {\bf 10}, 186 (1974).

\bibitem{Teaney:2003kp} 
  D.~Teaney,
  Phys.\ Rev.\ C {\bf 68}, 034913 (2003).
   
\bibitem{Baier:2006um} 
  R.~Baier, P.~Romatschke and U.~A.~Wiedemann,
  Phys.\ Rev.\ C {\bf 73}, 064903 (2006).

\bibitem{Monnai:2009ad} 
  A.~Monnai and T.~Hirano,
  Phys.\ Rev.\ C {\bf 80}, 054906 (2009).

\bibitem{Noronha-Hostler:2013gga} 
  J.~Noronha-Hostler, G.~S.~Denicol, J.~Noronha, R.~P.~G.~Andrade and F.~Grassi,
  Phys.\ Rev.\ C {\bf 88}, 044916 (2013).

\bibitem{Werner:2007bf} 
  K.~Werner,
  Phys.\ Rev.\ Lett.\  {\bf 98}, 152301 (2007).

\bibitem{Abelev:2013vea} 
  B.~Abelev {\it et al.}  [ALICE Collaboration],
  Phys.\ Rev.\ C {\bf 88}, no. 4, 044910 (2013).

\bibitem{Shen:2011eg} 
  C.~Shen, U.~Heinz, P.~Huovinen and H.~Song,
  Phys.\ Rev.\ C {\bf 84}, 044903 (2011).

\bibitem{Abelev:2012wca} 
  B.~Abelev {\it et al.}  [ALICE Collaboration],
  Phys.\ Rev.\ Lett.\  {\bf 109}, 252301 (2012)
  [arXiv:1208.1974 [hep-ex]].

\bibitem{Vredevoogd:2008id} 
  J.~Vredevoogd and S.~Pratt,
  Phys.\ Rev.\ C {\bf 79}, 044915 (2009);
  and Nucl.\ Phys.\ {\bf A830}, 515c (2009).
    
\bibitem{Aad:2013xma} 
  G.~Aad {\it et al.}  [ATLAS Collaboration],
  JHEP {\bf 1311}, 183 (2013).

\bibitem{Bernhard:2015hxa} 
  J.~E.~Bernhard, P.~W.~Marcy, C.~E.~Coleman-Smith, S.~Huzurbazar, R.~L.~Wolpert and S.~A.~Bass,
  arXiv:1502.00339 [nucl-th].

\bibitem{lhs}
  M. Stein, {\it Large sample properties of simulations using Latin Hypercube Sampling}, 
  Technometrics {\bf 29}, 143-151 (1987).

\bibitem{lhsPackage}
  Rob~Carnell, 
  {\it{lhs:\ Latin\ Hypercube\ Samples}},
  \url{http://cran.r-project.org/web/packages/lhs/index.html}.  

\bibitem{Preghenella:2012eu} 
  R.~Preghenella [ALICE Collaboration],
  arXiv:1203.5904 [hep-ex].

\bibitem{ATLAS:2012at} 
  G.~Aad {\it et al.}  [ATLAS Collaboration],
  Phys.\ Rev.\ C {\bf 86}, 014907 (2012).

\bibitem{Song:2010aq} 
  H.~Song, S.~A.~Bass and U.~Heinz,
  Phys.\ Rev.\ C {\bf 83}, 024912 (2011)
  [arXiv:1012.0555 [nucl-th]].

\bibitem{Huovinen:2009yb} 
  P.~Huovinen and P.~Petreczky,
  Nucl.\ Phys.\ {\bf A837}, 26 (2010).

\bibitem{Shen:2010uy} 
  C.~Shen, U.~Heinz, P.~Huovinen and H.~Song,
  Phys.\ Rev.\ C {\bf 82}, 054904 (2010).

\bibitem{Song:2013qma} 
  H.~Song, S.~Bass and U.~Heinz,
  Phys.\ Rev.\ C {\bf 89}, 034919 (2014).

\bibitem{Qiu:2011hf} 
  Z.~Qiu, C.~Shen and U.~Heinz,
  Phys.\ Lett.\ B {\bf 707}, 151 (2012).

\bibitem{Pratt:2015zsa} 
  S.~Pratt, E.~Sangaline, P.~Sorensen and H.~Wang,
  arXiv:1501.04042 [nucl-th].

\bibitem{Kovtun:2004de} 
  P.~Kovtun, D.~T.~Son and A.~O.~Starinets,
  Phys.\ Rev.\ Lett.\  {\bf 94}, 111601 (2005).

\end{thebibliography}
\end{document}